\documentclass[10pt,dvipsnames,svgnames]{article}

\usepackage[left=20mm,right=20mm,top=20mm,bottom=25mm]{geometry}
\usepackage[utf8]{inputenc}
\usepackage[T1]{fontenc}

\usepackage{tikz}
\usetikzlibrary{external}
\usepackage{multirow}
\usepackage{booktabs}
\usepackage{colortbl}

\usepackage{amssymb}
\usepackage{amsmath}
\usepackage{amsthm}
\usepackage{nicefrac}
\usepackage{bbm}
\theoremstyle{definition}   
\usepackage{chngcntr}
\usepackage{apptools}
\usepackage{relsize}
\usepackage{mathtools}

\usepackage{pgfplots}
\usepgfplotslibrary{groupplots}
\usepackage{pgfplotstable}
\pgfplotsset{scaled y ticks=false}
\usepackage{subcaption}
\usepackage[font=small]{caption}
\usepackage{wrapfig}
\pgfplotsset{compat=1.18}
\usepgfplotslibrary{fillbetween}
\usetikzlibrary{patterns}
\usepgfplotslibrary{colormaps}
\usetikzlibrary{backgrounds}

\usetikzlibrary{shapes}
\usetikzlibrary{arrows.meta} 

\tikzset{>=latex} 

\usepackage[maxbibnames=20, maxcitenames=2, 
    natbib=true, style=numeric,
    isbn=false, url=false,giveninits=true,sortcites]{biblatex}   
\bibliography{bibliography}

\usepackage[german, main=english]{babel}
\usepackage[autostyle]{csquotes}
\usepackage[T1]{fontenc}
\usepackage{hyperref}
\usepackage[affil-it]{authblk}
\usepackage[capitalise]{cleveref}
\crefname{section}{Sect.}{Sect.}
\crefname{appendix}{App.}{App.}
\crefname{table}{Tab.}{Tab.}

\usepackage{soul}
\sethlcolor{CPSorange!30!white}

\usepackage{framed}

\newcommand{\ba}{\boldsymbol{a}}
\newcommand{\bb}{\boldsymbol{b}}

\newcommand{\bt}{\boldsymbol{t}}

\newcommand{\bw}{\boldsymbol{w}}
\newcommand{\bx}{\boldsymbol{x}}
\newcommand{\bA}{\boldsymbol{A}}

\newcommand{\bB}{\boldsymbol{B}}

\newcommand{\bC}{\boldsymbol{C}}

\newcommand{\bF}{\boldsymbol{F}}

\newcommand{\bH}{\boldsymbol{H}}
\newcommand{\bI}{\boldsymbol{I}}

\newcommand{\bP}{\boldsymbol{P}}
\newcommand{\bQ}{\boldsymbol{Q}}

\newcommand{\bX}{\boldsymbol{X}}

\newcommand{\tr}{\operatorname{tr}}
\newcommand{\cof}{\operatorname{cof}}

\newcommand{\SL}{\operatorname{SL}}
\newcommand{\TS}{\operatorname{T}_{\SL(3)}}

\newcommand{\SO}{\text{SO}}

\newcommand{\bcI}{{\boldsymbol{\mathcal{I}}}}

\newcommand{\bbR}{{\mathbb{R}}}

\newcommand{\Cross}{\mathbin{\tikz [x=1.4ex,y=1.4ex,line width=.25ex] \draw (0.1,0.1) -- (0.9,0.9) (0.1,0.9) -- (0.9,0.1);}}

\usepackage{stmaryrd}
\usepackage{cleveref}

\newcommand{\norm}[1]{\left\lVert#1\right\rVert}

\tikzset{
    ncbar angle/.initial=90,
    ncbar/.style={
        to path=(\tikztostart)
        -- ($(\tikztostart)!#1!\pgfkeysvalueof{/tikz/ncbar angle}:(\tikztotarget)$)
        -- ($(\tikztotarget)!($(\tikztostart)!#1!\pgfkeysvalueof{/tikz/ncbar angle}:(\tikztotarget)$)!\pgfkeysvalueof{/tikz/ncbar angle}:(\tikztostart)$)
        -- (\tikztotarget)
    },
    ncbar/.default=0.5cm,
}

\tikzset{square left brace/.style={ncbar=0.25cm}}
\tikzset{square right brace/.style={ncbar=-0.25cm}}
 
\definecolor{CPSgreen}{RGB}{22,164,138}
\definecolor{CPSlightblue}{RGB}{104,143,198}
\definecolor{CPSdarkblue}{RGB}{67,83,132}
\definecolor{CPSgrey}{RGB}{204, 204, 204}
\definecolor{CPSdarkgrey}{RGB}{175, 175, 175}
\definecolor{CPSorange}{RGB}{246,163,21}
\definecolor{CPSred}{RGB}{194,76,76}

\definecolor{MAXgrey}{RGB}{225,225,225}

\definecolor{blueR}{RGB}{0,114,189}
\definecolor{redR}{RGB}{206,59,20}

\tikzstyle{connect_conv}=[->,shorten <=5pt,shorten >=5pt,thick] 
\tikzstyle{connect_conv_corner}=[->, to path={|- (\tikztotarget)},shorten <=5pt,shorten >=5pt,thick] 
\tikzstyle{connect_conv_dashed}=[shorten <=5pt,shorten >=5pt,thick,dashed] 
\tikzstyle{node_conv}=[draw=black, rounded corners = 1,  thick]

\tikzstyle{node}=[thick,circle,draw=black,minimum size=2,inner sep=0.5,outer sep=0.6]
\tikzstyle{node icnn}=[color=CPSorange!10!black,node,draw=black,fill=CPSorange!25, text = black]
\tikzstyle{node in_out}=[CPSlightblue!10!black,draw=black,fill=CPSlightblue!20, text=black,rounded corners = 2]
\tikzstyle{node out}=[CPSlightblue!10!black,draw=black,fill=CPSlightblue!25, text=black, rounded corners = 2]
\tikzstyle{node inv_pot}=[CPSred!10!black,draw=CPSred!10!black,fill=CPSred!25, text=black, rounded corners = 2]
\tikzstyle{connect}=[thick,black] 
\tikzstyle{connect arrow}=[-{Latex[length=4,width=3.5]},thick,black,shorten <=0.5,shorten >=1]
\tikzstyle{connect arrow dashed}=[-{Latex[length=4,width=3.5]},ultra thick,black,shorten <=0.5,shorten >=1, 
dashed]

\tikzstyle{node_icnn}=[CPSorange!10!black,draw=CPSorange!10!black,fill=CPSorange!25, text=black, rounded corners = 2]

\tikzstyle{node NN}=[CPSorange!10!black,draw=CPSorange!10!black,fill=CPSorange!25, text=black, rounded corners = 2]

\newtheorem{definition}{Definition}[section]  
\newtheorem{remark}[definition]{Remark}

\title{
Neural networks meet hyperelasticity:
\\
A monotonic approach
\vspace{1ex}} 

\author[1,*]{Dominik~K.~Klein}
\author[2]{Mokarram~Hossain}
\author[3]{Konstantin~Kikinov}
\author[1]{\\Maximilian~Kannapinn}
\author[3]{Stephan~Rudykh}
\author[2]{Antonio~J.~Gil}
\affil[1]{\footnotesize Cyber-Physical Simulation, Department of Mechanical Engineering,\protect\\Technical University of Darmstadt, 64293 Darmstadt, Germany}
\affil[2]{\footnotesize Zienkiewicz Institute for Data, Modelling and AI, Faculty of Science and Engineering,\protect\\Swansea University, SA1 8EN, United Kingdom}
\affil[3]{School of Mathematical and Statistical Sciences, University of Galway, Galway, Ireland}
\affil[*]{\footnotesize Corresponding author, email: klein@cps.tu-darmstadt.de}

\date{January 5, 2025}

\begin{document}
\maketitle
\par\noindent\rule{\textwidth}{0.4pt}
\begin{abstract}
We apply physics-augmented neural network (PANN) constitutive models to experimental uniaxial tensile data of rubber-like materials whose behavior depends on manufacturing parameters. For this, we conduct experimental investigations on a 3D printed digital material at different mix ratios and consider several datasets from literature, including Ecoflex at different Shore hardness and a photocured 3D printing material at different grayscale values.
We introduce a parametrized hyperelastic PANN model which can represent material behavior at different manufacturing parameters. The proposed model fulfills common mechanical conditions of hyperelasticity. In addition, the hyperelastic potential of the proposed model is monotonic in isotropic isochoric strain invariants of the right Cauchy-Green tensor. In incompressible hyperelasticity, this is a relaxed version of the ellipticity (or rank-one convexity) condition. Using this relaxed ellipticity condition, the PANN model has enough flexibility to be applicable to a wide range of materials while having enough structure for a stable extrapolation outside the calibration data.
The monotonic PANN yields excellent results for all materials studied and can represent a wide range of largely varying qualitative and quantitative stress behavior. Although calibrated on uniaxial tensile data only, it leads to a stable numerical behavior of 3D finite element simulations. 
The findings of our work suggest that monotonicity could play a key role in the formulation of very general yet robust and stable constitutive models applicable to materials with highly nonlinear and parametrized behavior.

\end{abstract}
\vspace*{2ex}

\par\noindent\rule{\textwidth}{0.4pt}

\section{Introduction}


Most living organisms act as unified systems where various portions have different stiffnesses \cite{miserez2008}, e.g., muscles and bones.
Considering the efficiency of living organisms, engineering applications such as soft robotics \cite{Pelrine_2002,Chen_Review_SoftRobots,Guo_Liu_Liu_Leng_2021} and actuators \cite{Athinarayanarao2023,DE_review_2008,gu2017} can also benefit from the combination of multiple different material properties within a device. This goes along with challenges \cite{Lipson_2014} such as ensuring cohesion between the different applied materials. Recent advantages in material fabrication offer a promising alternative to create artificial systems mimicking living creatures \cite{bartlett2015}. In additive manufacturing, the properties of a single material can be varied, notably within one manufactured part \cite{Slesarenko_Rudykh_2018,zhang2024}. Using various 3D printing techniques, soft materials with spatially varied stiffness can be manufactured where resins with different properties can be combined in the printing process \cite{Slesarenko_Rudykh_2018,Arora_Li_Rudykh_2022}, or the degree of cure of a single photoactive base resin can be varied by applying different light intensities during the manufacturing process \cite{zhang2024,valizadeh2021}.
Apart from 3D printing, several commercially available soft materials (e.g., Ecoflex, Dragon skin, Sylgard) applicable for mold manufacturing appeared, which can be synthesized with base resins \cite{Liao_Hossain_Yao_2020} of various portions to create complex systems with varied stiffness. 
Overall, these materials offer promising applications in soft robotics \cite{Yue2023,Chen_Review_SoftRobots,bartlett2015}, metamaterials \cite{Li_Rudykh_2019,Zheng2024,Slesarenko_Rudykh_2016}, soft actuators \cite{Athinarayanarao2023,moreno2022}, or energy harvesters \cite{collins2021}, to mention but a few. 
Finally, the dependency of material properties on manufacturing parameters was also reported for different material classes, e.g., hydrogels \cite{Wang_Zhong_Xiao_Qu_2024}, liquid crystal elastomers \cite{WEI2025105977}, and stiff 3D printed thermoplastics \cite{hernandez2020}.

\medskip

To fully exploit the capabilities of such \emph{architected} materials, efficient and accurate simulation tools are required. In particular, constitutive models that represent their mechanical behavior. For this, conventional constitutive models such as the hyperelastic Mooney-Rivlin model with parametrized material parameters have been applied \cite{zhang2024,valizadeh2021,Liao_Hossain_Yao_2020}. By parametrizing the material parameters in quantities such as the grayscale value in digital light processing 3D printing \cite{valizadeh2021}, one single constitutive model can represent the material behavior for different manufacturing parameters \cite{zhang2024,valizadeh2021}. These conventional models have one major drawback. Even for materials without parametric dependencies, the choice and calibration of a suitable constitutive models out of the multitude available is challenging and requires a lot of expert knowledge \cite{Hossain_Steinmann_2013,Steinmann_Hossain_Possart_2012,Ricker_Wriggers_2023}. This gets even more complicated for architected materials due to their nonlinear dependency on manufacturing parameters \cite{valizadeh2023}. Since different architected materials can exhibit vastly different mechanical behavior, for conventional constitutive models, the challenging process of material modeling has to be repeated for each new type of material.

\medskip 


This is where machine learning comes into play. Recently, constitutive models based on physics-augmented neural networks (PANNs) have become well-established \cite{rosenkranz2024,tac2023,zlatic2023,linka2023,fuhg2024}. These models combine the flexibility that neural networks (NNs) offer \cite{Hornik1991} with a sound mechanical basis \cite{linden2023}.\footnote{The paradigm of combining machine learning methods with scientific knowledge is not restricted to material modeling but widespread in many scientific fields \cite{rueden2021,peng2021,karniadakis2021,Kumar2022,kannapinn}.} Due to their flexibility, PANN constitutive models do not have to be tailored to a specific material class, but can represent a lot of different materials with one unified approach \cite{tac2023,kalina2024b}.
In hyperelastic PANN constitutive modeling, NNs are applied to represent strain energy potentials \cite{Linka2020}. NN potentials can be formulated in terms of strain invariants \cite{kalina2022b,Linka2020} or in terms of the components of strain tensors \cite{Fernandez2020,klein2021,Vlassis2022a}. By complementing the NN potential by additional growth \cite{klein2021} and normalization terms \cite{linden2023}, all mechanical conditions of hyperelasticity can be fulfilled by construction \cite{linden2023}. Several PANN models with convexity properties were proposed, which was shown to improve the model's stability and generalisation \cite{klein2021,kalina2024a,asad2022}. Based on convex neural network architectures \cite{Amos2016}, PANN models can be formulated to be polyconvex \cite{klein2021} which ensures stability properties by construction. In \textcite{kalina2024a}, a relaxed version of polyconvexity is applied by approximately fulfilling polyconvexity through loss terms. In \cite{asad2022,Zheng_Kochmann_Kumar_2024}, a heuristically motivated convexity condition in the right Cauchy-Green tensor is applied. 
Furthermore, a variety of parametrized PANN constitutive models has been proposed \cite{Linka2020,fernandezMaterialModelingParametric2022,schommartz2024,Klein_Roth_Valizadeh_Weeger_2023,LECLEZIO2024102260,VIJAYAKUMARAN2024106015}.
PANN constitutive models were successfully applied to soft biological tissues \cite{tac2023,linka2023a} including materials with parametric dependencies \cite{Linka_Cavinato_Humphrey_Cyron_2022}, rubber-like materials \cite{tac2023}, and synthetic homogenisation data of microstructured materials \cite{kalina2022b,klein2021,Linka2020}.
However, to the best of the author's knowledge, they have not yet been applied to architected rubber-like materials, particularly concerning real experimental data obtained from a wide range of soft polymeric materials synthesised under different manufacturing conditions.

\medskip


In this work, we apply parametrised hyperelastic PANN constitutive models to experimental data of architected rubber-like materials whose behavior depends on manufacturing parameters. We consider uniaxial tension tests of different materials with highly varying qualitative and quantitative stress behavior. This includes new experimental investigations on a 3D printed digital material and several datasets from the literature. We introduce a PANN model for which the hyperelastic potential is monotonic\footnote{In this work, if not stated otherwise, ``monotonic`` refers to component-wise monotonically increasing functions, i.e., $\partial_{x_i}f(\bx)\geq 0$.} in isotropic isochoric strain invariants of the right Cauchy-Green tensor. We show that, in incompressible hyperelasticity, this monotonicity condition is a relaxed version of the ellipticity (or rank-one convexity) condition. Using this relaxed ellipticity condition, the PANN model has enough flexibility to be applicable to a wide range of materials, while having enough structure for a stable extrapolation outside the calibration data. The monotonic PANN yields excellent results for all materials studied. Even when only being calibrated on uniaxial tensile data, the monotonic PANN leads to a stable numerical behavior of 3D finite element simulations. Overall, the findings of our work suggest that monotonicity could play a key role in the formulation of very general yet robust and stable constitutive models applicable to materials with highly nonlinear and parametrized behavior. 
The outline of the work is as follows. In~\cref{sec:basics}, we introduce the fundamentals of hyperelasticity. In~\cref{sec:PANN}, we introduce the PANN constitutive models applied in this work. In~\cref{sec:app}, we apply the models to experimental datasets and conduct finite element analysis. This is followed by the conclusion in~\cref{sec:conc}.

\section{Fundamentals of hyperelasticity}\label{sec:basics} 

In this section, we introduce the constitutive conditions of parametrized hyperelasticity in \cref{sec:const_cond}, followed by the basics of elliptic invariant-based modeling in \cref{sec:inv}. In \cref{sec:mono}, we demonstrate how, in invariant-based incompressible hyperelasticity, monotonicity can be applied as a relaxed ellipticity (or rank-one convexity) condition. 

 
\subsection{Constitutive conditions}\label{sec:const_cond}

Let us consider the hyperelastic potential
\begin{equation}\label{eq:pot}
\widetilde{W}:\operatorname{SL}(3)\times \bbR^m\times\bbR\rightarrow\bbR\,,\qquad (\bF;\;\bt;\;\gamma)\mapsto {W}(\bF;\;\bt)-\gamma(J-1)\,,
\end{equation}
which corresponds to the strain energy density stored in a parametrized, perfect incompressible body \cite{Holzapfel2000}. Here, $\bF\in\text{SL}(3)$ denotes the deformation gradient, where $\text{SL}(3):=\big\{\bX \in\allowbreak \;\mathbb{R}^{3\times 3}\,\rvert\,\allowbreak \det \bX =1\big\}$ is the special linear group in $\bbR^3$. The parameter vector characterizing material properties is denoted by $\bt\in\bbR^m$, while $\gamma\in\bbR$ is a Lagrange multiplier ensuring $J=\det\bF =1$. Note that the constitutive model introduced later only represents ${W}$, while $\gamma$ is received from boundary conditions and balance equations.
With the first Piola-Kirchhoff stress defined as the gradient field
\begin{equation}\label{eq:PK1}
    \bP
    =\partial_{\bF}{W}(\bF;\;\bt)-\gamma J\bF^{-T}\,,
\end{equation}
\textbf{thermodynamic consistency} is ensured by construction. The potential is subject to the \textbf{stress normalisation} condition
\begin{equation}\label{eq:stress_norm}
   \partial_{\bF}{W}(\bF;\;\bt)\big\rvert_{\bF=\bI}=\boldsymbol{0}\qquad \forall\,\bt\in\bbR^m\,.
\end{equation}
Assuming isotropic material behavior, \textbf{material symmetry} and \textbf{objectivity} are formalised as
\begin{equation}\label{eq:obj_sym}
\begin{aligned}
        {W}(\bF;\;\bt)={W}(\bQ_1\bF\bQ_2^T;\;\bt) \qquad\forall\,(\bF,\bt)\in\operatorname{SL}(3)\times\bbR^m,\,\bQ_1,\bQ_2\in\operatorname{SO}(3)\,,
\end{aligned}
\end{equation}
where $\SO(3):=\big\{\bX \in\allowbreak \mathbb{R}^{3\times 3}\;\rvert\allowbreak \;\bX^T\bX=\bI,\;\det \bX =1\big\}$ is the special orthogonal group in $\bbR^3$. Note that in hyperelasticity, objectivity implies fulfillment of the balance of angular momentum \cite[Proposition~8.3.2]{Silhavy2014}. Thus, the latter does not have to be introduced as an additional constitutive condition in this work.
Further conditions are grounded in the concept of \textbf{convexity} \cite{horak2023,kruzik2019}. To foster understanding of convexity conditions in finite elasticity theory, we consider potentials of the form
\begin{equation}\label{eq:pot_ext}
    \mathcal{W}:\operatorname{SL}(3)\times\operatorname{SL}(3)\times\bbR_+\times\bbR^m\rightarrow\bbR\,,\qquad (\bF,\,\bH,\,J;\,\bt)\mapsto\mathcal{W}(\bF,\,\bH,\,J;\,\bt)\,,
\end{equation}
with $W(\bF;\;\bt)=\mathcal{W}(\bF,\,\bH,\,J;\,\bt)$, where the extended set of arguments includes the deformation gradient $\bF$, its cofactor $\bH=\cof\bF=(\det\bF)\bF^{-T}=\frac{1}{2}\bF\Cross\bF$, and its determinant $J=\det\bF$. For sufficiently smooth convex functions, the Hessian is p.s.d. \cite{Silhavy2014}, which yields the general convexity condition\footnote{Throughout this work, tensor compositions and contractions are denoted by $\left(\bA\,\bB\right)_{ij}=A_{ik}B_{kj}$, $\bA:\bB=A_{ij}B_{ij}$, and $\bA:\mathbb{A}:\bB=A_{ij}\mathbb{A}_{ijkl}B_{kl}$, respectively, with second order tensors $\bA$ and $\bB$ and fourth order tensors $\mathbb{A}$. The tensor cross product operator $\Cross$ is defined as $(\bA\Cross\bB)_{iI}=\mathcal{E}_{ijk}\mathcal{E}_{IJK}A_{jJ}B_{kK}$, where $\mathcal{E}_{ijk}$ is the third-order permutation tensor.}
\renewcommand\arraystretch{1.2}
\begin{equation}\label{eq:ellip_operator}
\begin{aligned}
\bA:d^2 _{\bF\bF}\mathcal{W}:\bA
=
\begin{bmatrix}
    \bA: \\
    (\bA\Cross\bF): \\
    \bA:\bH
    \end{bmatrix}
    [\mathbb{H}_{\mathcal{W}}]
\begin{bmatrix}
    :\bA \\
    :(\bA\Cross\bF) \\
    \bA:\bH
    \end{bmatrix}
    +\left(\partial_{\bH}\mathcal{W}+\partial_{J}\mathcal{W}\,\bF\right):(\bA\Cross\bA)
    \geq 0 \,,
\end{aligned}
\end{equation}
with the Hessian operator $[\mathbb{H}_{\mathcal{W}}]$ defined as
\renewcommand\arraystretch{1.4}
\begin{equation}\label{eq:hessian_w}
   [\mathbb{H}_{\mathcal{W}}]:=  \begin{bmatrix}
   \partial^2_{\bF\bF} \mathcal{W}&\partial^2_{\bF\bH}\mathcal{W} & \partial^2_{\bF J}\mathcal{W}\\
\partial^2_{\bH\bF}\mathcal{W} & \partial^2_{\bH\bH}\mathcal{W} & \partial^2_{\bH J}\mathcal{W}\\
\partial^2_{J\bF }\mathcal{W} & \partial^2_{J \bH }\mathcal{W} & \partial^2_{JJ}\mathcal{W}
\end{bmatrix}\,.
\end{equation}
A convexity condition commonly applied in constitutive modeling is \textbf{polyconvexity} \cite{Ebbing2010,neff2015}. Polyconvex potentials allow for a (non-unique) representation $W(\bF;\;\bt)=\mathcal{W}(\bF,\,\bH,\,J;\,\bt)$, where $\mathcal{W}$ is a convex function in $(\bF,\,\bH,\,J)$ \cite{Ghiba_Martin_Neff_2018,Mielke}. Polyconvexity is linked to existence theorems in finite elasticity theory \cite{Ball1976,Ball1977}. These existence theorems make assumptions on the hyperelastic potential far outside a practically relevant deformation range \cite{kruzik2019}, raising questions about their practical relevance \cite{klein2021}. However, besides its relevance in existence theorems, polyconvexity has been recognised as the most straightforward way of ensuring ellipticity by construction \cite{neff2015}. The  \textbf{ellipticity} (or rank-one convexity) condition is linked to the concept of \textbf{material stability}, which ensures stability when applying the constitutive model in numerical simulations \cite{neff2015,Schroeder_Neff_Balzani_2005}.   
In incompressible hyperelasticity, the ellipticity condition is given by
\begin{equation}\label{eq:ellip}
\begin{aligned}
       (\ba\otimes\bB):d^2 _{\bF\bF}\mathcal{W}:(\ba\otimes\bB)\geq 0 \qquad \forall \ba,\bB\in\bbR^3\quad\text{with}\quad&\bF+\ba\otimes\bB\in\SL(3)\,.
    \end{aligned}
\end{equation}
As the constitutive model is defined on $\bF\in\SL(3)$, the test vectors $\ba,\bB$ must satisfy $\bF+\ba\otimes\bB\in\SL(3)$. This means the hyperelastic potential must only be convex along rank-one directions contained in the $\SL(3)$, which can be expressed as \cite[Sec.~5.2]{neff2015}
\begin{equation}\label{eq:tangent_space}
\bF+\ba\otimes\bB\in\SL(3)\quad \Leftrightarrow\quad \ba\otimes\bB\in\TS(\bF)\,,
\end{equation}
where $\TS(\bF):=\big\{\bX \in\allowbreak \;\mathbb{R}^{3\times 3}\,\rvert\,\allowbreak \bX :\bF^{-T}=0\big\}$ is the tangent space to $\SL(3)$ at $\bF$ \cite{Dunn2003}. As a consequence of \cref{eq:tangent_space}, it holds that $(\ba\otimes\bB):\bH=0$, which simplifies \cref{eq:hessian_w,eq:ellip_operator} for ellipticity to\footnote{Note that the first-order derivatives in \cref{eq:ellip_operator} vanish for ellipticity due to $(\ba\otimes\bB)\Cross(\ba\otimes\bB)=\boldsymbol{0}$ \cite{bonet2015}.}
\renewcommand\arraystretch{1.2}
\begin{equation}\label{eq:ellip_operator_reduced}
\begin{aligned}
 (\ba\otimes\bB):d^2 _{\bF\bF}\mathcal{W}:(\ba\otimes\bB)
=
\begin{bmatrix}
     (\ba\otimes\bB): \\
    ( (\ba\otimes\bB)\Cross\bF):
    \end{bmatrix}
    [\overline{\mathbb{H}}_{\mathcal{W}}]
\begin{bmatrix}
    : (\ba\otimes\bB) \\
    :( (\ba\otimes\bB)\Cross\bF) 
    \end{bmatrix}
    \geq 0 \quad \forall \ba,\bB\in\bbR^3,\,\ba\otimes\bB\in\TS(\bF)\,,
\end{aligned}
\end{equation}
with the reduced Hessian operator
\renewcommand\arraystretch{1.4}
\begin{equation}\label{eq:hessian_w_2}
   [\overline{\mathbb{H}}_{\mathcal{W}}]:=  \begin{bmatrix}
   \partial^2_{\bF\bF} \mathcal{W}&\partial^2_{\bF\bH}\mathcal{W} \\
\partial^2_{\bH\bF}\mathcal{W} & \partial^2_{\bH\bH}\mathcal{W} 
\end{bmatrix}\,.
\end{equation}
This means that for incompressible hyperelasticity, ellipticity is independent of the hyperelastic potential's functional relationship in $J$.  
In incompressible hyperelasticity, ellipticity is equivalent to \cite[Eq.~(1.49)]{Zee1983}
\begin{equation}\label{eq:ellip_conds}
       \begin{rcases}
\big(\bQ(\bF,\bB)\Cross\bQ(\bF,\bB)\big):\left(\bF^{-T}\bB\otimes\bF^{-T}\bB\right)\geq 0  \\
\big(\bQ(\bF,\bB)\Cross\bI\big):\left(\bF^{-T}\bB\otimes\bF^{-T}\bB\right)\geq 0
    \end{rcases}    \forall \bB\in\bbR^3\quad \text{with}\quad
    \big(\bQ(\bF,\bB)\big)_{ik}=\big(\partial^2_{\bF\bF}W\big)_{ijkl}B_jB_l\,,
\end{equation}
where $\bQ(\bF,\bB)$ is the acoustic tensor in direction $\bB$. 

\begin{remark}
In compressible hyperelasticity, ellipticity requires positive semi-definiteness of the acoustic tensor, which is equivalent to
\begin{equation}\label{eq:ellip_conds_comp}
       \begin{rcases}
\big(\bQ(\bF,\bB)\Cross\bQ(\bF,\bB)\big):\bQ(\bF,\bB)\geq 0  \\
\big(\bQ(\bF,\bB)\Cross\bQ(\bF,\bB)\big):\bI\geq 0 \\
\big(\bQ(\bF,\bB)\Cross\bI\big):\bI\geq 0
    \end{rcases}    \forall \bB\in\bbR^3\,,
\end{equation}
Thus, the two conditions obtained for incompressible hyperelasticity (cf.~\cref{eq:ellip_conds}) are a relaxed version of the three conditions obtained for compressible hyperelasticity (cf.~\cref{eq:ellip_conds_comp}). Again, this is a consequence of~\cref{eq:tangent_space}. 
\end{remark}

\subsection{Elliptic invariant-based modeling}\label{sec:inv}

By formulating the hyperelastic potential in terms of isochoric strain invariants of the right Cauchy-Green tensor $\bC=\bF^T\bF$, i.e., $W(\bF;\;\bt)=\psi(\bcI;\;\bt)$ with
\begin{equation}\label{eq:pot_inv}
{\psi}:\bbR^2_+\times\bbR^m\,,\qquad (\bcI;\;\bt)\mapsto\psi(\bcI;\;\bt)\,,
\end{equation}
and
\begin{equation}
\bcI=(\Bar{I}_1,\,\Bar{I}_2)\,,\qquad\Bar{I}_1=J^{-2/3}\tr\bC\,,\qquad \Bar{I}_2=J^{-4/3}\tr\cof\bC\,,
\end{equation}
stress normalisation, material symmetry, and objectivity (cf.~\cref{eq:obj_sym,eq:stress_norm}) can be fulfilled. With this choice of strain invariants, the Lagrange multiplier $\gamma$ in \cref{eq:pot} equals the hydrostatic pressure \cite{sansour2018}.
Considering the ellipticity condition \cref{eq:ellip_operator_reduced,eq:hessian_w_2}, in incompressible hyperelasticity, both $\Bar{I}_1$ and $\Bar{I}_2$ are clearly elliptic due to the p.s.d. of $\partial^2_{\bF\bF} \Bar{I}_1=\partial^2_{\bH\bH}\Bar{I}_2=2\mathbb{I}$, with the fourth-order identity tensor $\mathbb{I}$. 

\begin{remark}
In {compressible} hyperelasticity, $\Bar{I}_1$ is polyconvex (and thus elliptic), while $\Bar{I}_2$ is not elliptic (and thus not polyconvex) \cite{Hartmann2003}.  To address the non-ellipticity of $\Bar{I}_2$, in compressible hyperelasticity, the slightly adapted invariant $\Bar{I}_2^*=\Bar{I}_2^{1.5}$ can be applied, which is polyconvex (and thus elliptic) \cite{Hartmann2003}. This difference in the convexity properties of  $\Bar{I}_2$ between compressible and incompressible hyperelasticity is caused by the condition in \cref{eq:tangent_space}, which is only present in incompressible hyperelasticity and reduces the Hessian operator in the ellipticity condition from \cref{eq:hessian_w} to \cref{eq:hessian_w_2}.
\end{remark}

\subsection{Monotonicity as a relaxed ellipticity condition}\label{sec:mono}

For the invariant-based potential \cref{eq:pot_inv}, the Hessian operator \cref{eq:hessian_w_2} for the ellipticity condition \cref{eq:ellip_operator_reduced} can be expressed as
\renewcommand\arraystretch{2}
\begin{equation}\label{eq:hessian_inv}
[\overline{\mathbb{H}}_{\psi}]=\underbrace{4
\begin{bmatrix}
\partial^2_{\Bar{I}_1\Bar{I}_1}\psi\,\bF\otimes \bF  & \partial^2_{\Bar{I}_1\Bar{I}_2}\psi\,\bF\otimes\bH \\
\partial^2_{\Bar{I}_1\Bar{I}_2}\psi\,\bH\otimes\bF  & \partial^2_{\Bar{I}_2\Bar{I}_2}\psi\,\bH\otimes \bH  
 \end{bmatrix}
 }_{\text{constitutive type term}}
 +
 \underbrace{
 2
\begin{bmatrix}
\partial_{\Bar{I}_1}\psi\,\mathbb{I} & \boldsymbol{0}\\
\boldsymbol{0} & \partial_{\Bar{I}_2}\psi\,\mathbb{I} 
 \end{bmatrix}
 }_{\text{geometric type term}}\,.
\end{equation}
i.e., $[\overline{\mathbb{H}}_{\mathcal{W}}]=[\overline{\mathbb{H}}_{\psi}]$ in \cref{eq:ellip_operator_reduced}. Since the invariants $\Bar{I}_1,\,\Bar{I}_2$ are nonlinear functions of $(\bF,\,\bH)$, the Hessian operator $[\overline{\mathbb{H}}_{\psi}]$ consists of two terms. The first term includes second derivatives of the potential with respect to the invariants, which suggests to phrase it as a ``constitutive type term``. The second term in \cref{eq:hessian_inv} includes first derivatives of the potential with respect to the invariants, which suggests to phrase it as a ``geometric type term``. 
For the invariant-based potential \cref{eq:pot_inv} to preserve the ellipticity of the invariants $\Bar{I}_1$ and $\Bar{I}_2$, the Hessian operator $[\overline{\mathbb{H}}_{\psi}]$ must be p.s.d. In theory, negative eigenvalues of the constitutive type term can be compensated by positive eigenvalues of the geometric type term and vice versa, resulting in an overall p.s.d. Hessian operator $[\overline{\mathbb{H}}_{\psi}]$. 
In constitutive modeling practice, however, it is infeasible to allow for negative eigenvalues in one of the two terms and still ensure \emph{by construction} that $[\overline{\mathbb{H}}_{\psi}]$ is p.s.d. for all deformation scenarios. Rather,~p.s.d. of both the constitutive type term and the geometric type term is applied by formulating the potential $\psi$ as a {convex and monotonic} function in the invariants \cite{Schroeder2003,klein2021}. Clearly, this is a {sufficient, but not necessary} condition for $[\overline{\mathbb{H}}_{\psi}]$ to be p.s.d., and more restrictive than it would have to be.

\medskip

To arrive at less restrictive model formulations while still including rank-one convexity to some extent, we propose using hyperelastic potentials for which the geometric type term is p.s.d., while the constitutive type term remains generic. This is achieved by \textbf{monotonicity in the invariants}:
\begin{equation}\label{eq:mono_invs}
\partial_{\Bar{I}_i}\psi(\bcI;\;\bt)\geq 0 \qquad \forall \, (i,\bcI,\bt)\in\mathbb{N}_{\leq 2}\times \bbR^2_+\times\bbR^m\,.
\end{equation}
For potentials fulfilling \cref{eq:mono_invs}, the geometric type term in \cref{eq:hessian_inv} is p.s.d. Similarly, hyperelastic potentials for which the constitutive type term is p.s.d. while the geometric type term remains generic could be formulated, by considering potentials which are convex in $\bcI$. To anticipate the model application in \cref{sec:app}, monotonicity of the potential in the invariants seems to be more reasonable than convexity of the potential in the invariants. 
Note that neither approach ensures ellipticity of the hyperelastic potential by construction. However, in case of a loss of ellipticity, the Hessian operator \cref{eq:hessian_inv} can be stabilised. For this, different methods have been proposed. E.g., in \cite{lee2012}, the eigenvalues of the full Hessian operator are numerically calculated, and the negative ones are set to zero, while in \cite{poya2024}, closed-form representations for the eigenvalues of the constitutive and geometric type term are introduced, with which again the negative eigenvalues can be pruned.
Finally, in a heuristic fashion, we also consider \textbf{monotonicity in the parameters}:
\begin{equation}\label{eq:mono_param}
\partial_{t_i}\psi(\bcI;\;\bt)\geq 0  \qquad \forall \,(i,\bcI,\bt)\in \mathbb{N}_{\leq m}\times \bbR^2_+ \times \bbR^m\,,
\end{equation}
which seems to be a reasonable assumption for some material classes \cite{Klein_Roth_Valizadeh_Weeger_2023}. Note that \cref{eq:mono_param} does not imply monotonicity of the stress in the parameters, for which every component of the mixed second derivative $\partial^2_{\bF\bt}W(\bF;\;\bt)$ would have to be nonnegative. Model formulations fulfilling the latter could quickly become overly restrictive by resulting in potentials which are convex in the deformation gradient alone instead of the extended set of arguments of the polyconvexity condition (cf.~\cref{sec:const_cond}).

\begin{remark}
Besides its relevance in the Hessian operator \cref{eq:hessian_inv}, monotonicity of the potential in isochoric isotropic invariants of the right Cauchy-Green tensor also ensures fulfillment of the {Baker-Ericksen (B-E) inequalities} \cite{baker_ericksen}
\begin{equation}
    (\sigma_i-\sigma_j)(\lambda_i-\lambda_j)\geq 0\,,
\end{equation}
with $i\neq j$, and where $\sigma_i$ are the principal values of the Cauchy stress $\boldsymbol{\sigma}=\bF^T\bP$ and $\lambda_i$ are the corresponding principal values of the deformation gradient $\bF$. The B-E inequalities are one of the weakest constitutive inequalities in hyperelasticity and state that, for each deformation state, the larger principal Cauchy stress occurs in the direction of the larger principal stretch. 
While it seems reasonable, even intuitive, that the larger stress occurs in the direction of the larger strain, the question in \emph{which} stress measure and \emph{which} strain measure this condition is formulated is fundamental \cite{neff2015}. Corresponding conditions in different stress and strain measures have been shown to easily lead to unphysical models \cite{neff2015,rivlin2004}. In contrast, the B-E inequalities allow for a physically reasonable material behavior.
The B-E inequalities are fulfilled if the inequality $\partial_{\Bar{I}_1}\psi+\lambda_i^2\partial_{\Bar{I}_2}\psi\geq 0\,,\,\, i\in\{1,2,3\}$ holds \cite{baker_ericksen,Zee1983}. Thus, monotonicity in the invariants (cf.~\cref{eq:mono_invs}) is a sufficient but not necessary condition for the B-E inequalities.
\end{remark}
\begin{remark}
Similar constitutive models can formulated by considering hyperelastic potentials which are monotonic in different sets of isotropic invariants such as $(\tr\bC,\,\tr\bC^2)$, or even for anisotropic invariants \cite{kalina2024b}. This would go along with a different structure of the Hessian operator (cf.~\cref{eq:hessian_inv}) regarding its constitutive and geometric type term \cite{poya2024}. Furthermore, depending on the considered invariants, the B-E inequalities might not be fulfilled by construction. 
If the invariant-based potential \cref{eq:pot_inv} takes $J$ as an additional argument, no monotonicity conditions should be posed on the functional relationship of the potential in $J$. E.g., in the case of isotropic hyperelasticity, requiring monotonically increasing potentials in $I_1=\tr\bC,I_2=\tr\cof\bC$ and $J$ would be unphysical as this generally does not allow for unstressed reference states \cite[Sec.~3.3.1]{linden2023}, while monotonically increasing potentials in $I_1,I_2$ which are monotonically decreasing in $J$ would be unphysical as this would imply unphysical ordered-force inequalities \cite[Sec.~2.2]{neff2015}. 
\end{remark}

\section{Physics-augmented neural network constitutive models}\label{sec:PANN}

In this section, we discuss different parametrized hyperelastic constitutive models based on physics-augmented neural networks (PANNs) in \cref{sec:PANNs}, which are based on monotonic and convex neural networks introduced in \cref{sec:MNNs_ICNNs}.


\subsection{Monotonic and convex neural networks}\label{sec:MNNs_ICNNs}

In this work, monotonic neural networks (MNNs) \cite{Klein_Roth_Valizadeh_Weeger_2023} and convex-monotonic neural networks (CMNNs) \cite{Amos2016} based on fully-connected feed-forward neural networks (FFNNs) are applied \cite{kollmannsberger2021} to represent hyperelastic potentials. In a nutshell, FFNNs are multiple compositions of vector-valued functions, where the components are referred to as nodes or neurons, and the function acting in each node is referred to as activation function. With input $\bx^{(0)}\in\bbR^{n_0}$, output $\bx^{(H+1)}\in\bbR$, and $H$ hidden layers, we consider FFNNs given as the mapping
\begin{equation}\label{eq:MNN}
\bx^{(h)}=\sigma^{(h)}(\bw^{(h)}\bx^{(h-1)}+\bb^{(h)})\in\bbR^{n_h}\,,\qquad h=1,\,\dotsc H+1\,.
\end{equation}
Here, $\bw^{(h)}$ and $\bb^{(h)}$ denote weights and bias of the NN, which together form the set of parameters $\boldsymbol{\mathcal{P}}$ that are optimized to fit the model to a given dataset. The component-wise applied activation functions are denoted by $\sigma^{(h)}$. A FFNN is called a MNN if its output $\bx^{(H+1)}\in\bbR$ is monotonic in its input $\bx^{(0)}\in\bbR^{n_0}$. A FFNN is called a CMNN if its output $\bx^{(H+1)}\in\bbR$ is convex and monotonic in its input $\bx^{(0)}\in\bbR^{n_0}$. 
To lay the foundational intuition for the construction of MNNs, we consider the univariate composite function
\begin{equation}\label{eq:1d_f}
    f:\bbR\rightarrow\bbR\,,\quad x\mapsto f(x):=(g\circ h)(x)\,,
\end{equation}
with $g,h:\bbR\rightarrow\bbR$, and assume sufficient differentiability of $g$ and $h$. Here, we apply the compact notation $(g\circ h)(x)=g(h(x))$. The function $f$ is monotonic when its first derivative
\begin{equation}
    f'(x)=[g'\circ h(x)]\,h'(x)\geq 0
\end{equation}
is non-negative, which is fulfilled if both $f$ and $g$ are monotonic functions ($g'\geq 0,\,h'\geq 0$), cf.~\textcite[Sec.~2]{Klein_Roth_Valizadeh_Weeger_2023} for a visual example. Recursive application of this provides conditions for arbitrary many function compositions. When all functions are monotonic, the overall function is monotonic. This can be extended to monotonicity of vector-valued function compositions, where each function must be monotonic, cf.~\textcite{klein2021} for an explicit proof.
Essentially, FFNNs are composite functions, and these conditions can readily be applied to construct the mapping \cref{eq:MNN}  to be monotonic, i.e., to be a MNN \cite{Klein_Roth_Valizadeh_Weeger_2023}: (i) all weights $\bw^{(h)}$ are non-negative and (ii) all activation functions $\sigma^{(h)}$ are monotonic. If at least one activation function $\sigma^{(h)}$ is not convex, the mapping \cref{eq:MNN} is not convex.
In a similar fashion, conditions for the univariate composite function \cref{eq:1d_f} to be convex can be derived. The function $f$ is convex when its second derivative
\begin{equation}
    f''(x)=[g''\circ h(x)]\,h'(x)^2+[g'\circ h(x)]\,h''(x)\geq 0
\end{equation}
is non-negative. A sufficient (albeit not necessary) condition for this is that $g$ is a convex and monotonic function ($g''\geq 0,\,g'\geq 0$) and $h$ is a convex function ($h''\geq 0$). Generalisation of this to convexity of vector-valued function compositions provides sufficient conditions for the mapping \cref{eq:MNN} to be convex \cite{Amos2016}: (i) the weights $\bw^{(h)}$ are non-negative in all layers besides the first one, (ii) the activation functions $\sigma^{(h)}$ in the first hidden layer are convex, and (iii) the activation functions $\sigma^{(h)}$ in every following layer are convex and monotonic. Combining these conditions with the ones obtained for the construction of MNNs provides conditions for the construction of NNs which are both convex and monotonic, i.e., CMNNs: (i) the weights $\bw^{(h)}$ are non-negative in all layers, and (ii) the activation functions $\sigma^{(h)}$ are convex and monotonic in all hidden layers.


\subsection{PANN models based on different NN architectures}\label{sec:PANNs}

\begin{figure}[t!]
   \centering
   \begin{minipage}{.6\textwidth}
   \begin{subfigure}[b]{\textwidth}
    \centering
\resizebox{\textwidth}{!}{
\tikzsetnextfilename{PANN}

\begin{tikzpicture}[x=1.6cm,y=1.1cm]
  \large
  \def\shiftX{0.65}
  \def\shiftY{0}
  \def\NC{6} 
  \def\nstyle{int(\lay<\Nnodlen?(\lay<\NC?min(2,\lay):3):4)} 
  \tikzset{ 
    node 1/.style={node in_out},
    node 2/.style={node inv_pot},
    node 3/.style={node icnn},
    node 4/.style={node NN},
  }
  
\draw[white,fill=white](0.3,-1.2) rectangle++ (10,2.2);

\draw[CPSorange!10!black,draw=CPSorange!10!black,fill=CPSorange!25, text=black, rounded corners = 4](3.25+\shiftX,-1.2) rectangle++ (0.9,2.1);
  
 \node[node 1, outer sep=0.6] (1-1) at (0.5+\shiftX,0.5+\shiftY) {$\bF$};
 \node[node 1, outer sep=0.6] (1-2) at (0.5+\shiftX,-0.8+\shiftY) {$\bt$};
 
\node[node 1, outer sep=0.6] (2-1) at (1.9+\shiftX,0.5+\shiftY) {$\bcI=\big({\Bar{I}_1},{\Bar{I}_2}\big)$};
\node[] (2-a) at (1.5+\shiftX, 0.5){};
\node[] (2-b) at (2.5+\shiftX, 0.5){};

\node[] (3-1) at (3.7+\shiftX,-0.15+\shiftY) {NN};
\node[] (3-a) at (3.3+\shiftX, 0.5){};
\node[] (3-b) at (3.3+\shiftX,- 0.8){};
\node[] (3-c) at (4.1+\shiftX,- 0.15){};

\draw[connect arrow] (1-1) -- (2-1);
\draw[connect arrow] (1-2) -- (3-b);

\draw[connect arrow] (2-1) -- (3-a);

\draw[CPSred!60,fill=CPSred,fill opacity=0.02,rounded corners=4](5+\shiftX-0.38+0.2,0.1-0.65) rectangle++ (3.1,0.8);


\node[node 2, outer sep=0.6] (4-1) at (5+\shiftX+0.2,-0.15+\shiftY) {$\psi^{\text{NN}}_{\square}$};

\node (4-2) at (5+\shiftX+1.7,-0.15+\shiftY) {$-\gamma(J-1)=\psi^{\text{PANN}}_{\square}$};

\node (4-3) at (5+\shiftX+2.025+0.3+1.5-0.975,-0.15+\shiftY) {};

\node (4-1a) at (5+\shiftX-0.2,-0.15+\shiftY) {};
\draw[connect arrow] (3-c) -- (4-1a);

\node[node 1, outer sep=0.6] (6-1) at (6.5+\shiftX+3.5-0.975,-0.15+\shiftY) {$ \bP$};
\draw[connect arrow]  (4-3) -- (6-1);

\node[] at (5.7+\shiftX+3.6-0.975,0.8-0.65+\shiftY) {$\partial_{\bF}$};

  \end{tikzpicture}}     \caption{General PANN structure.}
    \end{subfigure}

    \vspace{0.7cm}
        
    \begin{subfigure}[b]{\textwidth}
    \centering
\resizebox{\textwidth}{!}{
\tikzsetnextfilename{NN_architectures}

\begin{tikzpicture}[x=1.6cm,y=1.1cm]
  \large
  \def\shiftX{3.48}
  \def\shiftY{0}
  \def\NC{6} 
  \def\nstyle{int(\lay<\Nnodlen?(\lay<\NC?min(2,\lay):3):4)} 
  \tikzset{ 
    node 1/.style={node in_out},
    node 2/.style={node inv_pot},
    node 3/.style={node icnn},
    node 4/.style={node NN},
  }
  

\draw[white,fill=white](0.3,-1.2) rectangle++ (10,2.2);


\node[node 1, outer sep=0.6] (1-1) at (0.5,0.5) {$\bcI$};
\node[node 1, outer sep=0.6] (1-2) at (0.5,-0.8) {$\bt$};

\node[node 4,outer sep=0.6] (3-1) at (1.75,0.5) {CMNN};
\node[node 4,outer sep=0.6] (3-2) at (1.75,-0.8) {MNN};

\node[node 2, outer sep=0.6] (4-1) at (3.1,-0.15) {$\psi^{\text{NN}}_{\mathfrak{cm}}$};

\draw[connect arrow] (1-1) -- (3-1);
\draw[connect arrow] (1-2) -- (3-2);

\draw[connect arrow] (3-2) -- (3-1);

\draw[connect arrow] (3-1) -| pic[xshift=3.1-1.75]{} (4-1);


\node[node 1, outer sep=0.6] (1-1) at (0.8+\shiftX,0.5) {$\bcI$};
\node[node 1, outer sep=0.6] (1-2) at (0.8+\shiftX,-0.8) {$\bt$};

\node[node 4,outer sep=0.6] (3-1) at (1.75+\shiftX,-0.15) {MNN};

\node[node 2, outer sep=0.6] (4-1) at (3+\shiftX,-0.15) {$\psi^{\text{NN}}_{\mathfrak{m}}$};

\draw[connect arrow] (1-1) -| pic[xshift=1.25]{} (3-1);
\draw[connect arrow] (1-2) -| pic[xshift=1.25]{} (3-1);

\draw[connect arrow] (3-1) -- (4-1);


\node[node 1, outer sep=0.6] (1-1) at (0.8+\shiftX*2,0.5) {$\bcI$};
\node[node 1, outer sep=0.6] (1-2) at (0.8+\shiftX*2,-0.8) {$\bt$};

\node[node 4,outer sep=0.6] (3-1) at (1.75+\shiftX*2,-0.15) {FFNN};

\node[node 2, outer sep=0.6] (4-1) at (3+\shiftX*2,-0.15) {$\psi^{\text{NN}}_{\mathfrak{u}}$};

\draw[connect arrow] (1-1) -| pic[xshift=1.25]{} (3-1);
\draw[connect arrow] (1-2) -| pic[xshift=1.25]{} (3-1);

\draw[connect arrow] (3-1) -- (4-1);

\end{tikzpicture}}  
\caption{NN architectures based on convex-monotonic neural networks (CMNNs), monotonic neural networks (MNNs), and feed-forward neural networks (FFNNs).}
    \end{subfigure}%
    \end{minipage}
     \hspace{0.015\textwidth}
   \begin{minipage}{.36\textwidth}
 \begin{subfigure}[b]{\textwidth}
       \centering
\resizebox{\textwidth}{!}{
\tikzsetnextfilename{PANN_comp}

\newcommand\dX{3}

\begin{tikzpicture}

\node at (3.35,-2) [right,text width=3cm,align=center] {thermodynamic consistency};
\node at (3.35,-3.5) [right,text width=3cm,align=center] {objectivity, isotropy, stress normalisation};
\node at (3.35,-5) [right,text width=3cm,align=center] {incompressibility};
\node at (3.35,-6.5) [right,text width=3cm,align=center] {{monotonicity} in the parameters};
\node at (3.35,-8) [right,text width=3cm,align=center] {monotonicity in the invariants};
\node at (3.35,-9.5) [right,text width=3cm,align=center] {{convexity} in the invariants};

\foreach \i [evaluate={\y=-2.75-\i*1.5;}] in {0,1,2,3,4}{ 
\draw[CPSgrey,thick] (11,\y) -- (3.5,\y);
}

\foreach \i [evaluate={\x=6.5+\i*1.5;}] in {1,2}{ 
 \draw[CPSgrey,thick] (\x,-0.5) -- (\x,-8.75-1.5);
 }
 

\node at (6.5+0.75,-0.8) [] {{$\psi_{\mathfrak{cm}}^{\text{PANN}}$}};
\node at (6.5+0.75+1.5,-0.8) [] {{$\psi_{\mathfrak{m}}^{\text{PANN}}$}};
\node at (6.5+0.75+3,-0.8) [] {{$\psi_{\mathfrak{u}}^{\text{PANN}}$}};

\draw[CPSgreen,fill=CPSgreen,fill opacity=0.03,rounded corners=2,thick](7-0.5,-2.75) rectangle++ (4.5,1.5);
\node at (8+0.75,-2) [thick,align=center,fill=white,draw=CPSgreen,rounded corners=2] {stress is a gradient field};

\draw[CPSgreen,fill=CPSgreen,fill opacity=0.03,rounded corners=2,thick](7-0.5,-4.25) rectangle++ (4.5,1.5);
\node at (8.75,-3.5) [thick,align=center,fill=white,draw=CPSgreen,rounded corners=2] {isochoric invariants};

\draw[CPSgreen,fill=CPSgreen,fill opacity=0.03,rounded corners=2,thick](7-0.5,-4.25-1.5) rectangle++ (4.5,1.5);
\node at (8.75,-3.5-1.5) [thick,align=center,fill=white,draw=CPSgreen,rounded corners=2] {Lagrange multiplier};

\draw[CPSgreen,fill=CPSgreen,fill opacity=0.03,rounded corners=2,thick](9-0.5-0.5,-8.75-1.5) -- (9-0.5-0.5,-8.75+1.5-1.5) -- (11-0.5-1,-8.75+1.5-1.5)  -- (11-0.5-1,-8.75+3) -- (7-0.5,-8.75+3) -- (7-0.5,-8.75-1.5) -- cycle;
\draw[CPSorange,thick,pattern=north west lines, pattern color=CPSorange,rounded corners=2,fill opacity=0.5](11-1.5,-5.75) -- (11,-5.75)  --  (11,-8.75-1.5)--  (9-1,-8.75-1.5) -- (9-1,-8.75+1.5-1.5)-- (9+2-1.5,-8.75+1.5-1.5) -- cycle;

\draw [CPSdarkblue, thick] (3.5,-7.25+1.5) to [square left brace] (3.5,-2.75+1.5);
\draw [CPSdarkblue, thick] (3.5,-8.75-1.5) to [square left brace] (3.5,-5.75);

\node [rotate=90,below,CPSdarkblue] at (2.8,-3.5) {General PANN structure};
\node [rotate=90,below,CPSdarkblue] at (2.8,-7.25-0.75) {NN architectures};

\draw [CPSdarkblue, thick] (11,-8.75-1.5) to [square right brace] (11,-5.75-1.5);

\node [rotate=90,CPSdarkblue] at (11.5,-7.25-1.5) {Sufficient for ellipticity};

\end{tikzpicture}}  
\caption{Constitutive conditions. Shaded conditions are not included in the model formulation, while the rest is fulfilled by construction.}
    \end{subfigure}%
   \end{minipage}
\caption{Illustration of the PANN constitutive models. The \textbf{convex and monotonic PANN} is denoted by \textbf{$\psi_{\mathfrak{cm}}^{\text{PANN}}$}, the \textbf{monotonic PANN} is denoted by \textbf{$\psi_{\mathfrak{m}}^{\text{PANN}}$}, and the \textbf{unrestricted PANN} is denoted by \textbf{$\psi_{\mathfrak{u}}^{\text{PANN}}$}.}
\label{fig:PANN}
\end{figure}
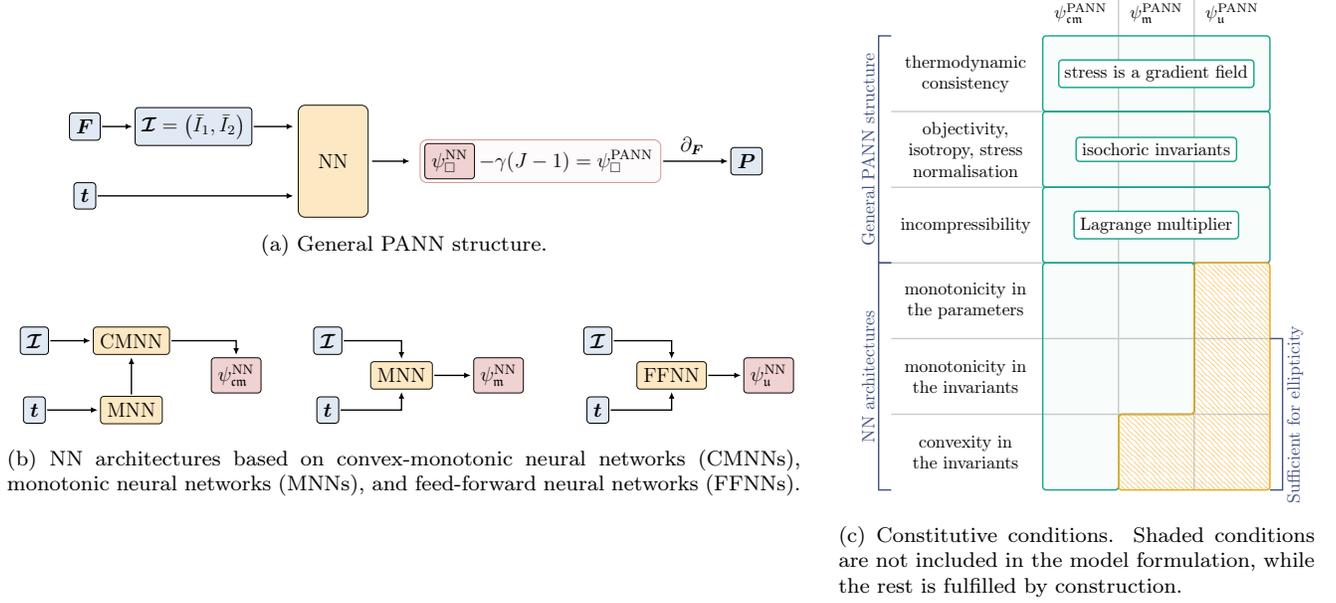

We consider hyperelastic potentials which depend on the invariants $\bcI=(\Bar{I}_1,\,\Bar{I}_2)\in\bbR^2_+$ and an additional parameter vector $\bt\in\bbR^m$, cf.~\cref{sec:inv}. The overall PANN model given by
\begin{equation}
\psi^{\text{PANN}}_{\square}=\psi^{\text{NN}}_{\square}(\bcI;\,\bt)-\gamma(J-1)\,
\end{equation}
consists of a NN potential $\psi^{\text{NN}}_{\square}$, where $\square\in\{\mathfrak{cm},\mathfrak{m},\mathfrak{u}\}$ denotes different NN architectures, and an additional term with the Lagrange multiplier $\gamma$ ensuring incompressibility (cf.~\cref{sec:const_cond}). With this general PANN structure, the model ensures thermodynamic consistency by defining the stress as a gradient field (cf.~\cref{eq:PK1}), while stress normalisation, material symmetry, and objectivity are ensured by formulating the potential in terms of isochoric invariants (cf.~\cref{sec:inv}). 
In this work, we consider three different NN architectures with different convexity and monotonicity properties. We apply a convex and monotonic NN potential which is denoted by $\psi^{\text{NN}}_{\mathfrak{cm}}$, a monotonic NN potential which is denoted by $\psi^{\text{NN}}_{\mathfrak{m}}$, and an unrestricted NN potential which is denoted by $\psi^{\text{NN}}_{\mathfrak{u}}$. The overall flow and structure of the PANN constitutive model is visualized in \cref{fig:PANN}, as well as an overview over the constitutive conditions the different models fulfill.

\paragraph{Convex and monotonic PANN model:} By formulating the hyperelastic potential as a convex and monotonic function of the invariants $\bcI=(\Bar{I}_1,\,\Bar{I}_2)$, ellipticity (or rank-one convexity) can be ensured by construction (cf.~\cref{sec:mono}). At the same time, the potentials functional dependency in the parameter vector $\bt$ should be monotonic but not convex in order to avoid an overly restrictive model formulation (cf.~\cref{eq:mono_param}). Both can be achieved by partially-input convex NNs \cite{Amos2016}. We apply the NN architecture
\begin{equation}\label{eq:NN_pot_cm}
\begin{aligned}
\bx^{(1)}=\mathcal{TH}(\bw^{(1)}\cdot\bt+\bb^{(1)})\quad &\in\bbR^{n}
\\
\bx^{(2)}=\mathcal{SP}(\bw^{(2)}\cdot(\Bar{I}_1-3,\Bar{I}_2-3,\bx^{(1)})^T+\bb^{(2)})\quad &\in\bbR^{n}\,,
\\
\psi^{\text{NN}}_{\mathfrak{cm}}=\bw^{(3)}\cdot \bx^{(2)}\quad &\in\bbR\,,
\end{aligned}
\end{equation}
with two hidden layers as proposed by \textcite{Klein_Roth_Valizadeh_Weeger_2023}, where the number of nodes in each hidden layer is $n$. Here, $\mathcal{TH}(x)=(e^{2x}-1)/(e^{2x}+1)$ denotes the hyperbolic tangent activation function, $\mathcal{SP}(x)=\log(1+e^x)$ denotes the softplus activation function, and a linear activation function $\mathcal{LIN}(x)=x$ is applied in the output layer. The hyperbolic tangent function is monotonic, while the softplus function is monotonic and convex. Given that all weights $\bw^{(h)}$ in \cref{eq:NN_pot_cm} are non-negative, the potential is convex in the invariants $\bcI$ and monotonic in $(\bcI,\,\bt)$ (cf.~\cref{sec:MNNs_ICNNs}). 

\paragraph{Monotonic PANN model:} In \cref{sec:mono}, we demonstrate how monotonicity can be applied as a relaxed ellipticity condition. For this, we consider hyperelastic potentials which are monotonic functions in $(\Bar{I}_1,\,\Bar{I}_2,\,\bt)$, and apply the NN architecture
\begin{equation}\label{eq:NN_pot_m}
\begin{aligned}
   \bx^{(1)}=\mathcal{TH}(\bw^{(1)}\cdot(\Bar{I}_1-3,\Bar{I}_2-3,\bt)^T+\bb^{(1)})\quad &\in\bbR^{n}\,,
\\
\bx^{(2)}=\mathcal{SP}(\bw^{(2)}\cdot\bx^{(1)}+\bb^{(2)})\quad &\in\bbR^{n}\,,
\\
\psi^{\text{NN}}_{\mathfrak{m}}=\bw^{(3)}\cdot\bx^{(2)}\quad &\in\bbR\,,
\end{aligned}
\end{equation}
with two hidden layers, where the number of nodes in each hidden layer is $n$. Given that all weights $\bw^{(h)}$ in \cref{eq:NN_pot_m} are non-negative, the resulting potential is monotonic in $(\Bar{I}_1,\,\Bar{I}_2;\;\bt)$ (cf.~\cref{sec:MNNs_ICNNs}). 

\paragraph{Unrestricted PANN model:} When the potential is not subject to convexity or monotonicity conditions, the choice of FFNN is generic. In this case, we apply NN architectures with one and two hidden layers and unrestricted weights, i.e., 
\begin{equation}\label{eq:NN_pot_u_1}
\begin{aligned}
\bx^{(1)}=\mathcal{TH}(\bw^{(1)}\cdot(\Bar{I}_1-3,\Bar{I}_2-3,\bt)^T+\bb^{(1)})\quad &\in\bbR^{n}\,,
\\
\psi^{\text{NN}}_{\mathfrak{u,1}}=\bw^{(2)}\cdot\bx^{(1)}\quad &\in\bbR\,,
\end{aligned}
\end{equation}
and
\begin{equation}\label{eq:NN_pot_u_2}
\begin{aligned}
   \bx^{(1)}=\mathcal{TH}(\bw^{(1)}\cdot(\Bar{I}_1-3,\Bar{I}_2-3,\bt)^T+\bb^{(1)})\quad &\in\bbR^{n}\,,
\\
\bx^{(2)}=\mathcal{SP}(\bw^{(2)}\cdot\bx^{(1)}+\bb^{(2)})\quad &\in\bbR^{n}\,,
\\
\psi^{\text{NN}}_{\mathfrak{u,2}}=\bw^{(3)}\cdot\bx^{(2)}\quad &\in\bbR\,,
\end{aligned}
\end{equation}
where the number of nodes in each hidden layer is $n$. The NN potentials in \cref{eq:NN_pot_cm,eq:NN_pot_m,eq:NN_pot_u_2} have $n^2+n(m+5)$ parameters, while the NN potential in \cref{eq:NN_pot_u_1} has $n(m+4)$ parameters.

\begin{remark}
Setting aside the nomenclature of machine learning, \cref{eq:NN_pot_cm,eq:NN_pot_m,eq:NN_pot_u_1,eq:NN_pot_u_2} are nothing more but mathematical functions suitable to be used for the representation of hyperelastic potentials. First of all, the NN potentials allow for a strong interrelation of the potential's input $(\Bar{I}_1,\Bar{I}_2,\bt)$, which is in contrast to many classical constitutive models which often consider additively decomposed functions in $\Bar{I}_1$ and $\Bar{I}_2$ \cite{Steinmann_Hossain_Possart_2012}, or even only consider $\Bar{I}_1$ \cite{kuhl2024a}. Furthermore, the flexibility of the NN potentials can be immediately increased, basically to an arbitrary amount \cite{Hornik1991}. This can be done by increasing the number of nodes or hidden layers. Note that for the convex and monotonic NN architectures introduced in \cref{eq:NN_pot_cm,eq:NN_pot_m}, the number of nodes and hidden layers could be increased by an arbitrary amount while still preserving convexity and monotonicity properties, given that the conditions introduced in \cref{sec:MNNs_ICNNs} are fulfilled \cite{klein2021}. In general, conventional constitutive models can also be formulated to be very flexible, e.g., by using a polynomial Mooney-Rivlin or Ogden models with a very large number of terms. However, the calibration of polynomials with a large number of terms eventually becomes infeasible, e.g., when very large or very small exponents are considered. In contrast to that, the calibration of PANN potentials has proven to be very stable, even for a large number of parameters \cite{klein2021,klein2024a}.
Finally, for shallow NN architectures as used in the present work, it is actually feasible to calculate the derivatives of the potential required for mechanical applications very efficiently in an analytical fashion \cite[Sec.~3.3]{franke2023}.
\end{remark}

\section{Application to experimental data}\label{sec:app}

In this section, we apply the PANN constitutive models discussed in \cref{sec:PANN} to a wide range of uniaxial tensile tests obtained from various soft polymers. In \cref{sec:datasets}, we introduce details on the model calibration and the considered datasets. This is followed by a NN hyperparameter study in \cref{sec:hyperparam}. In \cref{sec:uniax}, we apply the PANN models to the experimental datasets. Finally, in \cref{sec:stab}, we investigate the PANN model performance in multiaxial deformation scenarios.

\subsection{Model calibration and considered datasets}\label{sec:datasets}

 \renewcommand{\arraystretch}{1.15}
\begin{table}[t]
 \centering
  \caption{Number of calibration datapoints and calibration epochs for the different studies.}
 \begin{tabular}
 {lccccccccc}
\toprule
study &(0) &(I) &(II) &(III) &(IV) &(V) &(VI)  &(VII)\\
  \cmidrule(lr ){2-9} 
\# of calibration datapoints &87 &87 &56 &51 &42 &40 &30  &72\\
\# of calibration epochs &4e4 &3e4 &3e4 &2e4 &2e4 &2e4 &4e4  &3e4\\
  \bottomrule
 \end{tabular}
 \label{tab:hyperparam}
\end{table}

All materials considered in this work are characterized by a single parameter normalized to be in a unit interval for the calibration data, i.e., $t\in[0,1]\in\bbR$. Furthermore, all materials are experimentally investigated by uniaxial tensile tests. This results in datasets of the form
\begin{equation} 
\begin{aligned}
        \mathcal{D}=\Big\{\left({^1\lambda},\, {^1P},\,{^1t}\right),\dotsc,
        \left({^m\lambda},\,{^mP},\,{^m t}\right)\Big\}\,,
\end{aligned}
\end{equation}
consisting of $m$ stretch-stress-parameter tuples. Here, $\lambda$ and $P$ denote stretch and first Piola-Kirchhoff stress in tensile direction.
The overall dataset is split into a calibration dataset consisting of $m_c$ datapoints and a test dataset consisting of $m_t=m-m_c$ datapoints. 
To fit the PANN model to a given dataset, the mean squared error
\begin{equation}
\begin{aligned}
        \mathcal{MSE}(\boldsymbol{\mathcal{P}})=\frac{1}{m_c}\sum_{i=1}^{m_c}&\norm{^iP-{^iP}^{\text{model}}_{\text{1ax}}({^i\lambda};\,\boldsymbol{\mathcal{P}})
        }^2\quad \text{with}\quad     P^{\text{model}}_{\text{1ax}}=2\left(\partial_{\Bar{I}_1}\psi^{\text{NN}}_{\square}
+\lambda^{-1}\partial_{\Bar{I}_2}\psi^{\text{NN}}_{\square}\right)\left(\lambda-\lambda^{-2}\right)
\end{aligned}
\end{equation}
is minimized, see e.g.~\cite{Steinmann_Hossain_Possart_2012} for a derivation of $P^{\text{model}}_{\text{1ax}}$. The PANN model is calibrated through its gradients, i.e., the stresses, which is referred to as Sobolev training \cite{vlassis2021}.
The models are implemented and calibrated in TensorFlow 2.10.0 using Python 3.9.13. For the parameter optimization, the stochastic ADAM optimizer with a learning rate of $2\cdot 10^{-3}$ is applied, with the full batch of training data, TensorFlow's default batch size, and no loss weighting. The number of calibration datapoints and calibration epochs varies between the different studies, see~\cref{tab:hyperparam}. In each study, the models are calibrated multiple times for each dataset to account for the randomly initialized parameters of the NN and the stochastic nature of the ADAM optimizer. The considered materials are now briefly introduced case by case. 

\begin{figure}[tp!]
\centering
\resizebox{0.95\textwidth}{!}{
\tikzsetnextfilename{hyperparam}

\newcommand\NMAX{10}
\newcommand\NALPHA{0.35}

\begin{tikzpicture}
\begin{groupplot}[
	group style = {group size = 3 by 1, vertical sep = 0.06*\textwidth,
								horizontal sep = 0.15*\textwidth},
	]

 \nextgroupplot[
xlabel={stretch},
ylabel = {stress in MPa}, 
width=0.4*\textwidth, height=0.350*\textwidth,
ymin = 0, ymax = 6,
xmin = 1, xmax = 2.9,
xtick={1,2.5},
ytick={0,0,6},
yticklabels={\textcolor{white}{1.4},0,6},
xlabel shift = -14 pt,
ylabel shift = -19 pt,
title={\textbf{(a)}},
] 
    
\addplot [CPSdarkblue!100!CPSlightblue,mark=none,dashed,ultra thick,opacity=\NALPHA] table [x expr=\thisrowno{0}+1] {figure_code/DM_exp_data/9995-1.txt};
\addplot [CPSdarkblue!100!CPSlightblue,mark=none,dashed,ultra thick,opacity=\NALPHA] table [x expr=\thisrowno{0}+1] {figure_code/DM_exp_data/9995-2.txt};
\addplot [CPSdarkblue!100!CPSlightblue,mark=none, dashed,ultra thick] table [x expr=\thisrowno{0}+1] {figure_code/DM_exp_data/9995-3.txt};

\addplot [CPSdarkblue!50!CPSlightblue,mark=none,dashed,ultra thick,opacity=\NALPHA] table [x expr=\thisrowno{0}+1] {figure_code/DM_exp_data/9985-1.txt};
\addplot [CPSdarkblue!50!CPSlightblue,mark=none,dashed,ultra thick,opacity=\NALPHA] table [x expr=\thisrowno{0}+1] {figure_code/DM_exp_data/9985-2.txt};
\addplot [CPSdarkblue!50!CPSlightblue,mark=none,dashed,ultra thick] table [x expr=\thisrowno{0}+1] {figure_code/DM_exp_data/9985-3.txt};

\addplot [CPSdarkblue!0!CPSlightblue,mark=none,dashed,ultra thick,opacity=\NALPHA] table [x expr=\thisrowno{0}+1] {figure_code/DM_exp_data/9950-1.txt};
\addplot [CPSdarkblue!0!CPSlightblue,mark=none,dashed,ultra thick,opacity=\NALPHA] table [x expr=\thisrowno{0}+1] {figure_code/DM_exp_data/9950-2.txt};
\addplot [CPSdarkblue!0!CPSlightblue,mark=none,dashed,ultra thick] table [x expr=\thisrowno{0}+1] {figure_code/DM_exp_data/9950-3.txt};

\addplot [CPSred!0!CPSorange,mark=none,dashed,ultra thick,opacity=\NALPHA] table [x expr=\thisrowno{0}+1] {figure_code/DM_exp_data/elastico-1.txt};
\addplot [CPSred!0!CPSorange,mark=none,dashed,ultra thick,opacity=\NALPHA] table [x expr=\thisrowno{0}+1] {figure_code/DM_exp_data/elastico-2.txt};
\addplot [CPSred!0!CPSorange,mark=none,dashed,ultra thick] table [x expr=\thisrowno{0}+1] {figure_code/DM_exp_data/elastico-3.txt};

\node at (1.9,4.6) {\textcolor{CPSdarkgrey}{DM95}};   
\node at (1.9,2.55) {\textcolor{CPSdarkgrey}{DM85}};   
\node at (1.9,1.15) {\textcolor{CPSdarkgrey}{DM50}};   
\node at (2.65,1.15) {\textcolor{CPSdarkgrey}{Elastico}};   

\nextgroupplot[
legend columns = 1,
legend pos = north east,
xlabel={\# of nodes},
ylabel = {$\log_{10}$ MSE}, 
width=0.4*\textwidth, height=0.35*\textwidth,
ymin = -5, ymax = -2,
xmin = 1, xmax = 6,
xtick={1,2,3,4,5,6},
xticklabels={2,4,8,16,32,64},
ytick={-2,-3,-4,-5},
yticklabels={-2,-3,-4,-5},
title={\textbf{(b)}},
grid=major
] 

\addplot [CPSdarkblue, ultra thick, mark=none, name path = a] table [x index =0,y index=1]{figure_code/hyperparam/MSE_mono_2HL.txt};
\addplot [CPSdarkblue!100!CPSlightblue,  ultra thick, mark=none, densely dashed, name path = a] table [x index =0,y index=1]{figure_code/hyperparam/MSE_unr_2HL.txt};
\addplot [CPSdarkblue!100!CPSlightblue, ultra thick, mark=none, loosely dashed, name path = a] table [x index =0,y index=1]{figure_code/hyperparam/MSE_unr_1HL.txt}; 
    
\addlegendentry{monotonic 2HL};
\addlegendentry{unrestricted 2HL};
\addlegendentry{unrestricted 1HL};

\nextgroupplot[
legend pos = north west,
xlabel={\# of nodes},
ylabel = {\# of non-zero NN parameters}, 
width=0.4*\textwidth, height=0.35*\textwidth,
ymin = 3, ymax = 12.750,
xmin = 1, xmax = 6,
xtick={1,2,3,4,5,6},
xticklabels={2,4,8,16,32,64},
ytick={4,6,8,10,12},
yticklabels={16,64,256,1024,4096},
title={\textbf{(c)}},
grid=major
] 
 
\addplot [CPSdarkblue, ultra thick, mark=none, name path = a] table [x index =0,y index=1]{figure_code/hyperparam/params_mono_2HL.txt};
\addplot [CPSdarkblue!100!CPSlightblue,  ultra thick, mark=none, densely dashed] table [x index =0,y index=1]{figure_code/hyperparam/params_unr_2HL.txt};
\addplot [CPSdarkblue!100!CPSlightblue, ultra thick, mark=none, loosely dashed, name path = a] table [x index =0,y index=1]{figure_code/hyperparam/params_unr_1HL.txt};

\addlegendentry{monotonic 2HL};
\addlegendentry{unrestricted 2HL};
\addlegendentry{unrestricted 1HL};

\end{groupplot}


\end{tikzpicture}
}
\caption{\textbf{(a):} Experimental investigations on a 3D printed digital material (DM) using soft Elastico polymer and stiff VeroWhite polymer as base materials. The DM material is manufactured for three different mix ratios. \textbf{(b):} Hyperparameter study for the PANN constitutive model. MSE for different number of nodes and hidden layers (HL). \textbf{(c):} Number of non-zero NN parameters for different number of nodes and hidden layers.} 
\label{fig:hyperparam}
\end{figure}
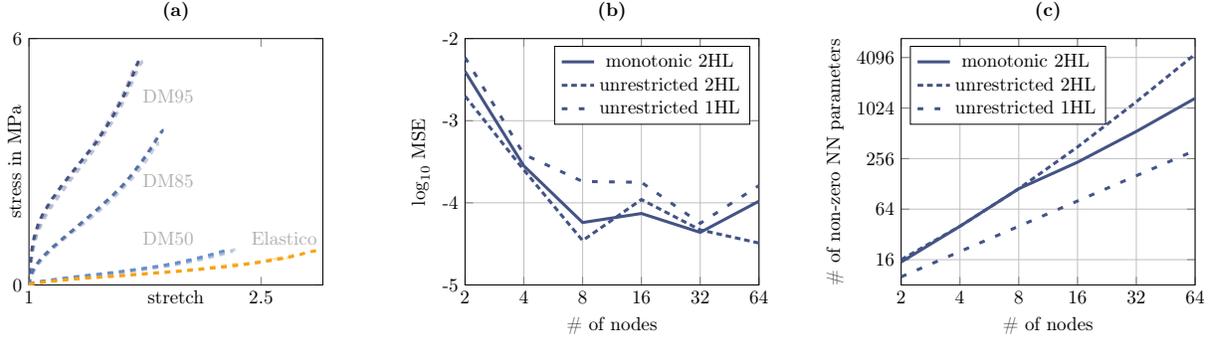

\paragraph{Digital Material (DM):} In polyjet 3D printing, small droplets of photoactive resins are applied on a printing platform and then cured by applying ultraviolet light. Multiple inkjet printing heads with different base resins can be combined to vary the final material behavior. For instance, when mixing soft TangoPlus polymer or soft Elastico polymer with stiff VeroWhite polymer, the so-called digital materials (DMs) that have mechanical stiffness of varied scales can be received. Thereby, different instances of DMs are characterized by their Shore A hardness scale. All of these materials are commercially available from Stratasys. 
We conducted experimental investigations on a DM printed with Elastico and VeroWhite. For this, we manufactured uniaxial tensile samples of type D412 at three different mix ratios using a Stratasys J35 3D printer. The material for different mix ratios is denoted by DM$\square$, where $\square\in\{50, 85, 95\}$ is the Shore A hardness value of the material. The specimens showed no significant dependence on the printing orientation in preliminary studies. We conducted quasi-static uniaxial tensile tests with a strain rate of $1.2\cdot 10^{-3}$ using an Instron 68SC-5 testing system, and determined tensile strain and strain rate in the gauge part of the samples with the help of a video-extensometer. Three specimens are used for each material type to ensure the reproducibility of the experiments. The stress curves for the different material types are visualized in \cref{fig:hyperparam}(a).\footnote{Throughout this work, the first Piola-Kirchhhoff stress is used for visualizations.} The DM50 material behaves mostly linear, while the DMs with larger Shore A hardness become increasingly nonlinear. With increasing Shore A value, the material's stiffness increases and its stretchability decreases. Apparently, the stress curves for each respective material type lie close to each other, suggesting reproducibility of the experimental setup. Thus, one single stress curve for each DM type is chosen. Note that the stress curve of the base material Elastico is not considered in the following investigations, and for the DM50 material, data until a stretch of $\lambda= 2$ is considered. In addition, we consider experimental data for DMs manufactured with TangoPlus and VeroWhite from \textcite{Slesarenko_Rudykh_2018}, presented in Fig.~2 therein (strain rate $1.2\cdot 10^{-3}$). We denote this material by DM-L$\square$, where $\square\in\{40,50,60,70, 85\}$ is the Shore A hardness value of the material for different mix ratios.

\paragraph{Digital Light Processing (DLP):} In grayscale DLP 3D printing, a photoactive resin is cured by exposing it to ultraviolet light. By varying the grayscale value $G$, i.e., the light intensity, the curing degree of the final material can be varied. This means that the light intensity and its duration influence the mechanical properties of the finally cured material. We consider recent experimental data for DLP materials from \textcite{zhang2024}, which is presented in Sec.~S.1 therein. We denote this material as DLP$\square$, where $\square\in\{0.1,0.2,0.3,0.4,0.5,0.7\}$ is the grayscale value used in the 3D printing process.

\paragraph{Ecoflex:} Ecoflex is a commercially available silicone elastomer that is synthesized by combining two base resins, which, after curing, form a soft rubber-like material. In most cases, they are cured at room temperature. By using different base resins with varied fractions, the material properties (e.g. stiffness) of the fully cured polymers can be varied. It means polymers of varied stiffness can be manufactured with a unified synthesis technique (e.g. with a two-part approach) for various applications. Thereby, different instances of Ecoflex are characterized by their Shore 00 hardness. We consider experimental data for {Ecoflex} polymer from \textcite{Liao_Hossain_Yao_2020} with different base resins, which is presented in Fig.~2(a) therein. We denote this material by ECO$\square$, where $\square\in\{10,20,30,50\}$ is the Shore 00 hardness value of the material for different base resins.


\subsection{NN hyperparameter study}\label{sec:hyperparam}

We conduct a preliminary hyperparameter study to find suitable numbers of hidden layers and nodes for the NN architecture. We investigate a monotonic PANN with two hidden layers, an unrestricted PANN with a single hidden layer, and an unrestricted PANN with two hidden layers, with the activation functions and architectures as introduced in \cref{eq:NN_pot_m,eq:NN_pot_u_1,eq:NN_pot_u_2}. The number of nodes is varied between 2 and 64. Each PANN model is calibrated five times to the full DM dataset, and the model instance with the lowest calibration loss is used for the following investigations. The remaining calibration details are provided in~\cref{sec:datasets} and \cref{tab:hyperparam}(0). 
The MSE values for the different PANN architectures are visualized in \cref{fig:hyperparam}(b). For the following investigations, NN architectures with two hidden layers and eight nodes are applied for the monotonic and the unrestricted model, which showed to be sufficiently flexible but still having a moderate NN size. For the convex and monotonic model, the NN architecture with two hidden layers introduced in \cref{eq:NN_pot_cm} is applied with eight nodes.

Furthermore, we investigate the influence of monotonicity on the PANN model's sparsity, i.e., the number of parameters that take zero-values \cite{mcculloch2024}. For the PANN with a single hidden layer, the overall number of parameters depends linearly on the number of nodes, while for the PANNs with two hidden layers, the overall number of parameters depends quadratically on the number of nodes (cf.~\cref{sec:PANNs}). The number of non-zero parameters for the different architectures are visualized in~\cref{fig:hyperparam}(c). For the unrestricted PANN, all weights and biases take non-zero values in every investigated model instance. Consequently, the number of non-zero NN parameters grows linearly for the single-layered models and quadratically for the double-layered models. In contrast to that, for the monotonic models, a lot of weights take zero values, given that the overall NN architecture is sufficiently large. For example, for the monotonic PANN with 16 nodes in the hidden layers, only around 66\% of the parameters are non-zero, while for 64 nodes, only around 30\% of the parameters are non-zero. This sparsity has potential benefits in an efficient implementation of the PANN constitutive models, e.g., for finite element simulations, as multiplications with zero values do not have to be evaluated. 


\begin{figure}[tp!]
\centering
\resizebox{0.95\textwidth}{!}{
\input{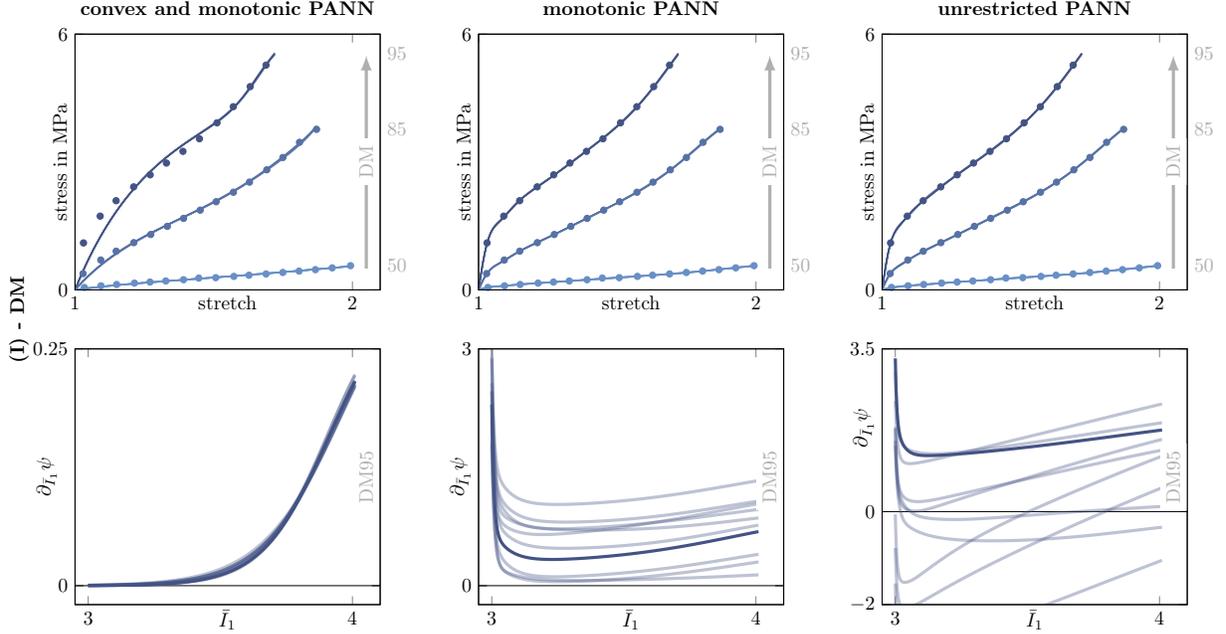}
}
\caption{Influence of monotonicity and convexity on the PANN model performance in uniaxial tension. Points denote the calibration data and lines denote the PANN models. \textbf{(I):} Interpolation of DM for different mix ratios.} 
\label{fig:stress_inv_1a}
\end{figure}

\subsection{Model performance in uniaxial tension}\label{sec:uniax}

We investigate the performance of different PANN constitutive models for the experimental data introduced in \cref{sec:datasets}. We apply a convex and monotonic PANN, a monotonic PANN, and an unrestricted PANN (cf.~\cref{sec:PANN}). In each investigation, the PANN models are calibrated 30 times, and the 10 models with the lowest calibration loss are used for evaluation. The remaining calibration details are provided in \cref{sec:datasets}.

\paragraph{(I) - Interpolation of the DM data:} We apply all three PANN models to the DM data. All datapoints are used for calibration, meaning we investigate interpolation in the deformation and parameter space. 
In \cref{fig:stress_inv_1a}(I), the stress prediction of the calibrated PANN models for different mix ratios are visualized. For the DM50 case, the stress behavior is mostly linear. When transitioning to DM85 and DM95, the DM material shows large quantitative and qualitative changes in the stress response and becomes increasingly nonlinear. Both the monotonic and the unrestricted PANN show excellent performance and have a close to perfect fit of the stress for all parameter values. The convex and monotonic PANN has an excellent interpolation for the DM50 case, while its performance decreases for DM85 and DM95. This can be explained by investigating the stress coefficients, i.e., the partial derivatives of the NN potential with respect to the strain invariants, which are also visualized in \cref{fig:stress_inv_1a}(I). For the unrestricted PANN, the functional relationship of the NN potential is generic. Thus, the stress coefficients can also take generic forms. For the monotonic PANN, the stress coefficients are non-negative (cf.~\cref{eq:mono_param}). While this is a restriction on the function space the model can represent, it is apparently an admissible restriction, as the stress prediction remains excellent. The convex and monotonic PANN further restricts the function space the model can represent. For this model, the stress coefficients are non-negative and seem to be monotonic.\footnote{Note that the stress coefficients $\partial_{\Bar{I}_1}\psi$ of the convex and monotonic model must not necessarily be monotonic functions, which would mean that $\partial^2_{\Bar{I}_1\Bar{I}_1}\psi\geq 0$. Rather, for the convex and monotonic  PANN, the second derivatives of the potential are restricted by the constitutive type term in \cref{eq:hessian_inv}, i.e., a convexity condition in $(\Bar{I}_1,\Bar{I}_2)$.} Apparently, this is overly restrictive for the investigated DM material, and consequently, the convex and monotonic PANN has a poor fit of the stress.

\begin{figure}[tp!]
\centering
\resizebox{0.95\textwidth}{!}{
\input{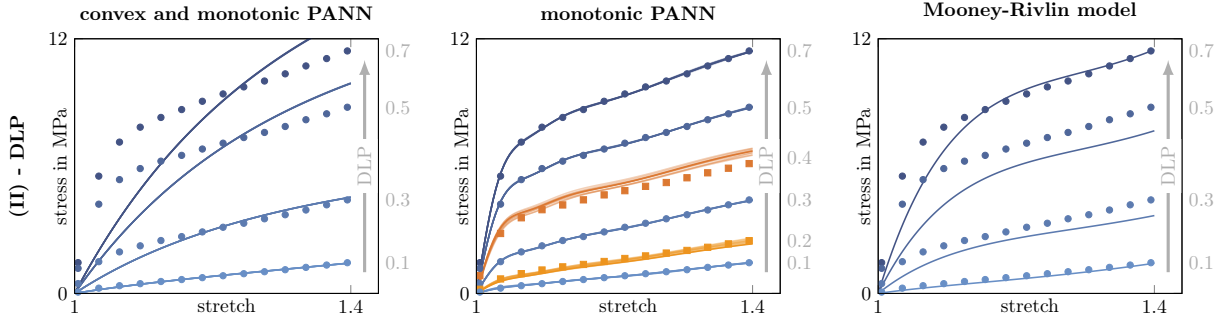}
}
\caption{Influence of monotonicity and convexity on the PANN model performance in uniaxial tension and comparison with a Mooney-Rivlin model from literature \cite{zhang2024}. Points denote the calibration data, squares denote the test data, and lines denote the PANN models. \textbf{(II):} Interpolation of DLP for different grayscale values.} 
\label{fig:stress_inv_1b}
\end{figure}

\paragraph{(II) - Interpolation of the DLP data:} We apply the convex and monotonic PANN, the monotonic PANN, and a conventional constitutive model from literature to the DLP data. In particular, we consider the Mooney-Rivlin model  
\begin{equation}\label{eq:MR_param}
    \psi^{\text{MR}}=c_{10}(G)(I_1-3)+c_{01}(G)(I_2-3)+c_{11}(G)(I_1-3)(I_2-3)
\end{equation}
applied in \textcite{zhang2024} to represent data for the DLP material. In \cite{zhang2024}, a parametrization of the material parameters in the grayscale value $G$ is applied, e.g., $c_{10}(G)=(114.3G^3-207.3G^2+23.99G-1.143)\,\text{MPa}$. Note that this Mooney-Rivlin model is not monotonic or convex. All datapoints for the DLP\{0.1,0.3,\allowbreak0.5,\allowbreak0.7\} data are used for calibration of the PANN models, which were also included in the calibration of \cref{eq:MR_param} by \textcite{zhang2024}. The DLP\{0.2, 0.4\} data is used as test dataset. This study is an interpolation in the deformation and parameter space. 
In \cref{fig:stress_inv_1b}(II), the stress prediction of the calibrated models for different grayscale values are visualized. For small grayscale values, the DLP material behaves mostly linear. For increasing grayscale values, the DLP material shows large quantitative and qualitative changes in the stress response. For large grayscale values, this includes a pronounced change of slope of the stress response around small stretch values. 
While the convex and monotonic PANN and the Mooney-Rivlin model have good predictions for DLP0.1, their performance gets worse for increasing grayscale values. Apparently, these models are not sufficiently flexible to represent the pronounced nonlinearity of the DLP material. Since these models fail even in interpolating the calibration data, the test data is not further investigated. In contrast to that, the monotonic PANN model shows excellent performance for all grayscale values, even for the ones not included in the calibration process. In this case, the unrestricted PANN (for which no results are visualized) performs equally well.

\begin{figure}[tp!]
\centering
\resizebox{0.95\textwidth}{!}{
\input{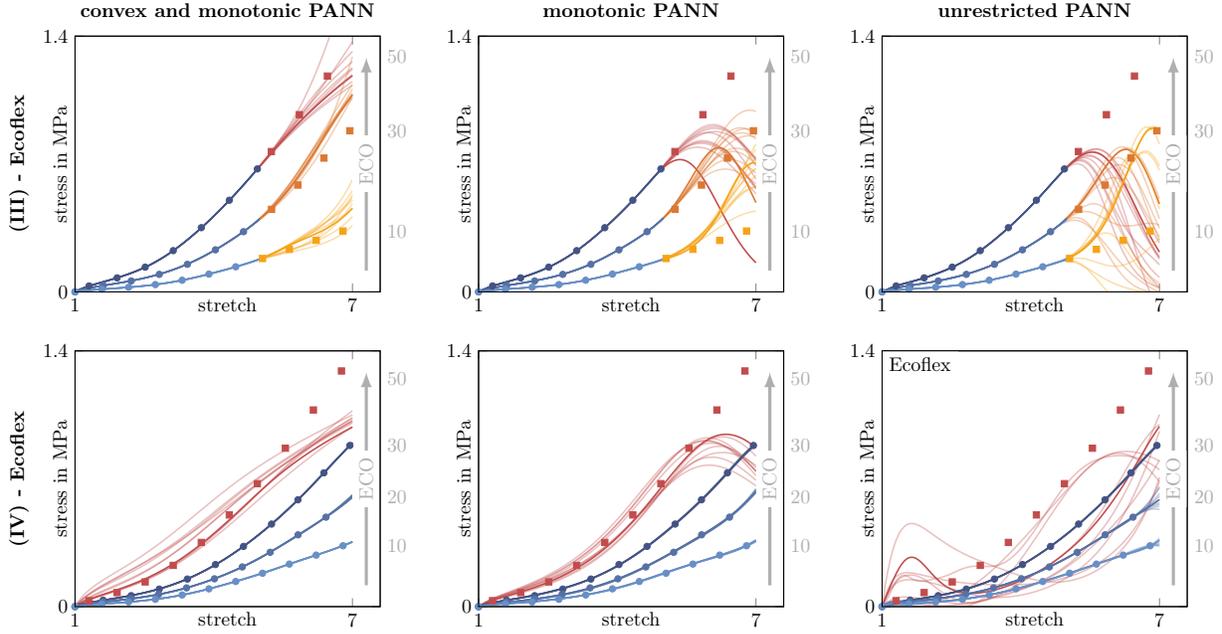}
}
\caption{Influence of monotonicity and convexity on the PANN model performance in uniaxial tension. Points denote the calibration data, squares denote the test data, and lines denote the PANN models. \textbf{(III):} Extrapolation in the strain space for Exoflex with different base resins, \textbf{(IV):} Extrapolation in the parameter space for Ecoflex with different base resins.} 
\label{fig:stress_inv_2}
\end{figure}

\paragraph{(III) - Extrapolation in the deformation space for the Ecoflex data:} We apply all three PANN models to the Ecoflex data. The ECO\{10, 30, 50\} data for stretches of up to 5 is used for calibration, while the data for stretch values between 5 and 7 is used as test dataset. This study investigates extrapolation in the deformation space and interpolation in the parameter space. 
In \cref{fig:stress_inv_2}(III), the stress prediction of the calibrated models for different base resins are visualized. On the calibration dataset, all models perform excellent. The convex and monotonic PANN performs very well for the test dataset, with only moderate deviations from the test data even for extrapolations of 200\% stretch. For the monotonic PANN, some model instances perform decent for moderate extrapolations. For more pronounced extrapolations, almost all model instances fail to predict the ground truth data and show a pronounced decrease in stress for an increase in stretch, which does not seem mechanically reasonable. The unrestricted PANN performs even worse in extrapolation, where most model instances fail even for small extrapolation. Again, for the unrestricted PANN, we observe a pronounced decrease in stress for an increase in stretch. The deviation between the different calibrated model instances is the smallest for the convex and monotonic PANN and the largest for the unrestricted PANN.

\paragraph{(IV) - Extrapolation in the parameter space for the Ecoflex data:} We apply all three PANN models to the Ecoflex data. The ECO\{10, 20, 30\} data is used for calibration, while ECO50 data is used as test dataset. Stretch values of up to 7 are used for calibration and testing. This study investigates interpolation in the deformation space and extrapolation in the parameter space.
In \cref{fig:stress_inv_2}(IV), the stress prediction of the calibrated models for different base resins are visualized. On the calibration dataset, all models perform excellent. The convex and monotonic PANN performs very well for the test dataset, with moderate deviations from the test data for larger stretch values. The monotonic PANN shows an excellent performance for the test data for up to moderate stretch values but bad predictions for larger stretch values. The unrestricted PANN fails to extrapolate even for small stretch values and shows an unphysical oscillating stress behavior for most model instances.

\begin{figure}[tp!]
\centering
\resizebox{0.95\textwidth}{!}{
\input{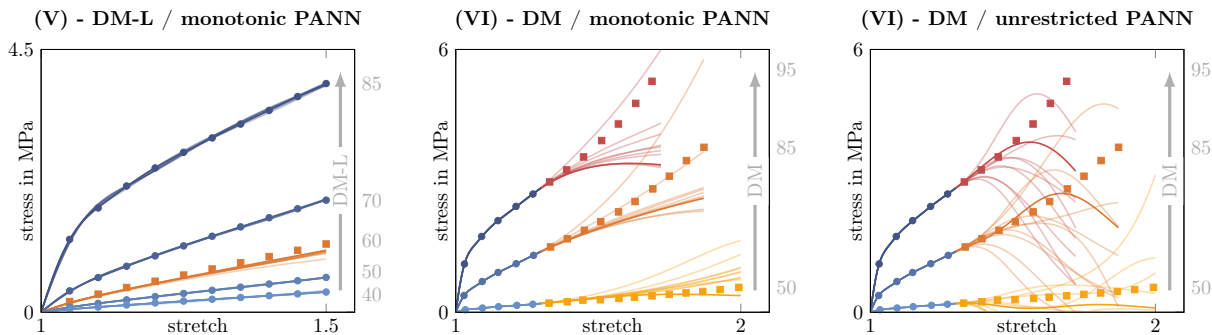}
}
\caption{Influence of monotonicity on the PANN model performance in uniaxial tension. Points denote the calibration data, squares denote the test data, and lines denote the PANN models. \textbf{(V):} Interpolation of DM-L for different mix ratios, \textbf{(VI):} Extrapolation in the strain space for DM for different mix ratios.} \label{fig:stress_inv_3}
\end{figure}

\paragraph{(V) - Interpolation of the DM-L data:} We apply the monotonic PANN to the DM-L data. The  DM-L\{40, 50, 70, 85\} data is used for calibration, while the DM-L60 data is used as test data. This study investigates interpolation in the deformation and parameter space. 
In \cref{fig:stress_inv_3}(V), the stress prediction of the calibrated models for different mix ratios are visualized. The monotonic PANN model performs excellent for both the calibration and test datasets. In this case, the unrestricted PANN (for which no results are visualized) performs equally well.

\paragraph{(VI) - Extrapolation in the deformation space for the DM data:}  We apply the monotonic and the unrestricted PANN to the DM data. The DM data for stretches of up to 1.3 are used for calibration, while the remaining stretch values are used as test dataset. This study investigates extrapolation in the deformation space and interpolation in the parameter space.
In \cref{fig:stress_inv_3}(VI), the stress prediction of the calibrated models for different mix ratios are visualized. On the calibration dataset, both models perform excellent. For the test dataset, the prediction of the monotonic PANN is decent. This is in contrast to the unrestricted PANN, which shows bad prediction qualities even for moderate extrapolations. The deviation between the different calibrated model instances is considerably smaller for the monotonic PANN than for the unrestricted PANN.

\paragraph{Discussion:} The monotonic PANN performs well in all cases studied, where two aspects are particularly noteworthy. First of all, for some materials, the stress magnitude between the lowest and the largest stress path is highly different. It means that for the DLP material, the maximum stress value for DLP0.7 is almost eight times as high as for DLP0.1. Still, the monotonic PANN model almost perfectly predicts both the load paths. Secondly, the shape of the stretch-stress curves not only highly varies between the different materials, e.g., the stress curves of Ecoflex are increasing for larger stretch values, while the stress curves of DLP show a plateau behaviour for larger stretch values. More than that, for materials such as DLP and DM, a large change of both the stress magnitude and the shape of the stress curve occurs for varying parameter values. With the monotonic PANN, a single modeling approach can capture all of these different material characteristics. In contrast to that, the convex and monotonic PANN apparently is too restrictive for some materials, where it has a bad fit even on the calibration data. 
The unrestricted and the monotonic PANN perform equally well on the calibration data and for interpolation in the parameter space. Where the convex and monotonic PANN is applicable, i.e., where it can interpolate the calibration data, all three models perform equally well on the calibration data. The model performance mainly deviates when extrapolation away from the calibration data is investigated. Generally, for the studies where the convex and monotonic PANN works out, it outperforms the remaining models in extrapolation scenarios. The monotonic PANN, in turn, outperforms the unrestricted model in extrapolation scenarios. Overall, this demonstrates how providing a model with more structure can improve its generalisation, but it can also reduce its flexibility and, thus, its applicability. 

On a critical note, the experimental data used in this work only contains uniaxial tension tests. However, the aim of constitutive modeling is to find a model that represents the behavior of a material for \emph{general} deformation modes, which is not restricted to uniaxial tension. With the data at hand, we can't assess the prediction quality of our models for general deformation scenarios. That being said, the observation that for some materials, PANNs that are convex and monotonic are too restrictive is well-known from PANN applications including multiaxial deformation modes \cite{klein2024a,kalina2024a}. The same goes for the observation that including convexity and monotonicity in PANN models can improve their generalisation \cite{kalina2024a,asad2022,kalina2024b}.
Furthermore, conventional constitutive models such as Carroll's model \cite{carroll2011} are monotonic in strain invariants and have been successfully applied to multiaxial experimental data. This suggests that our monotonic PANN model is also applicable to multiaxial data, at least for some materials.


\subsection{Model performance for multiaxial deformation}\label{sec:stab}

While we cannot assess the prediction quality of our models for general deformation modes, we can still assess if their predictions are mechanically reasonable, i.e., if they incline with the constitutive conditions introduced in \cref{sec:const_cond}. The monotonic PANN model fulfills most of these conditions such as thermodynamic consistency by construction (cf.~\cref{sec:PANN}). The only relevant condition that our monotonic PANN does not fulfill by construction is ellipticity (cf.~\cref{eq:ellip}). This condition ensures material stability, i.e., a good convergence behavior when applying the constitutive model in numerical simulation methods such as the finite element method \cite{Schroeder_Neff_Balzani_2005,Ebbing2010}. In this section, it is demonstrated that monotonicity can improve the model's material stability and, thus, its applicability in multiaxial finite element analysis (FEA).
Monotonic and unrestricted PANNs are calibrated to the DM data for the following investigations. The DM data for all mix ratios and for stretch values of up to 1.7 is used for calibration. The models are calibrated 30 times, and the 5 models with the best calibration loss are investigated. The remaining calibration details are provided in~\cref{sec:datasets} and \cref{tab:hyperparam}(VII). For the following FEA, the PANN models were implemented in COMSOL Multiphysics.

\begin{figure}[tp!]
\centering
\resizebox{0.8\textwidth}{!}{
\input{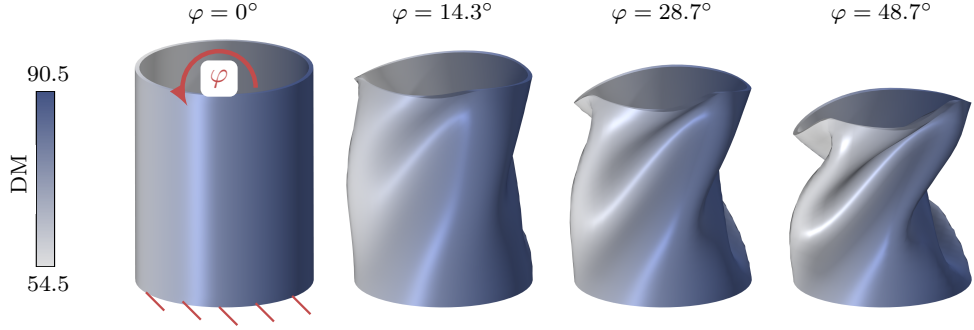}
}
\caption{Simulation of a cylinder. The cylinder is clamped on one end while a torsional displacement is applied on the other. Simulation results for a monotonic PANN calibrated to the DM data.}
\label{fig:FEA_cylinder}
\end{figure}

\begin{figure}[tp!]
\centering
\resizebox{0.95\textwidth}{!}{
\input{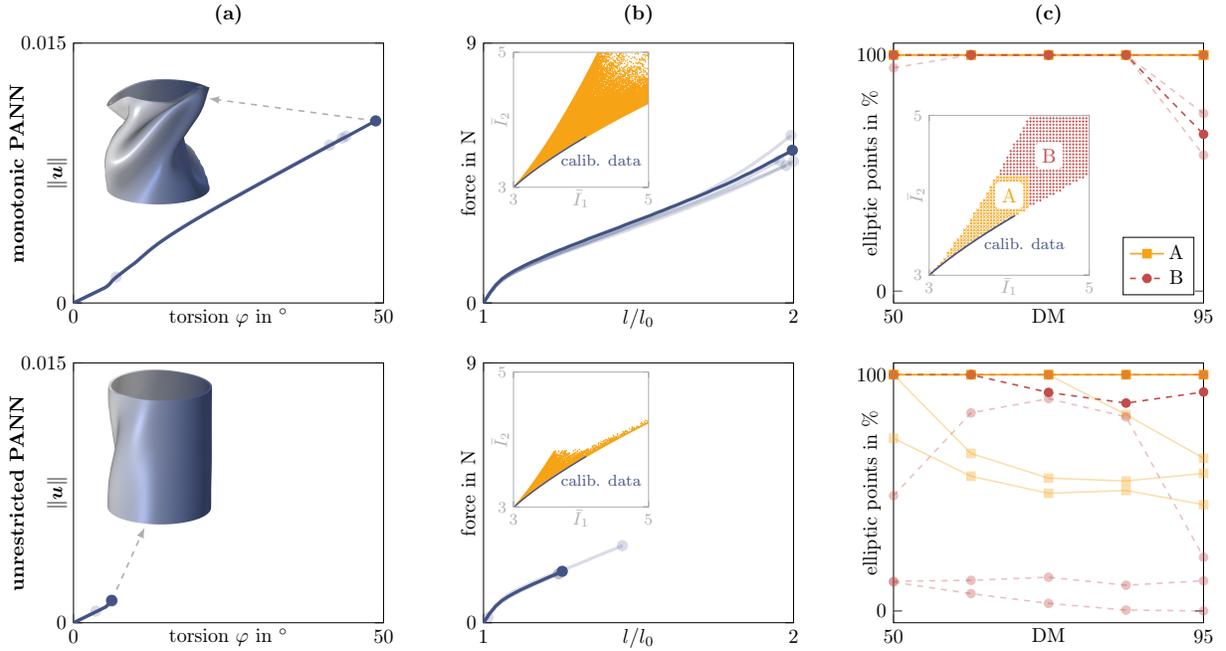}
}
\caption{Influence of monotonicity on the PANN model performance in multiaxial FEA. \textbf{(a):} $L^2$ norm of displacement of the cylinder under torsion. \textbf{(b):} Resulting tensile force of the biaxial specimen. \textbf{(c):} Ellipticity of the calibrated PANNs for general deformation scenarios.
}
\label{fig:mat_stab_1}
\end{figure}

\paragraph{Finite-element analysis of a cylinder:} At first, we demonstrate the applicability of the monotonic PANN constitutive model in FEA including numerically challenging buckling phenomena. For that, we simulate torsion of a cylinder with a linear grading between DM54.5 and DM90.5 (cf.~\cref{fig:FEA_cylinder}). The cylinder is clamped on one end, while a torsional displacement is applied on the other. The cylinder has an inner and outer radius of $r_{\text{min}}=0.48\text{cm}$ and $r_{\text{max}}=0.5\text{cm}$, respectively, and a length of $l=2r_{\text{max}}$. The simulation result for the best monotonic PANN is visualized in~\cref{fig:FEA_cylinder}. The cylinder experiences a pronounced buckling, which is numerically challenging and demonstrates stability of the monotonic PANN when applied in complex FEA. Furthermore, in \cref{fig:mat_stab_1}(a), the $L^2$ norm of the displacement for the simulations with the monotonic and the unrestricted PANN are visualized up to the point where the simulation does not converge anymore. Four of the five investigated monotonic PANNs converge for torsions up to over 40$^{\circ}$. For the unrestricted PANN, the simulations only converge for moderate torsions, where most simulations stop converging at around 6$^{\circ}$ and are not able to simulate pronounced buckling. 

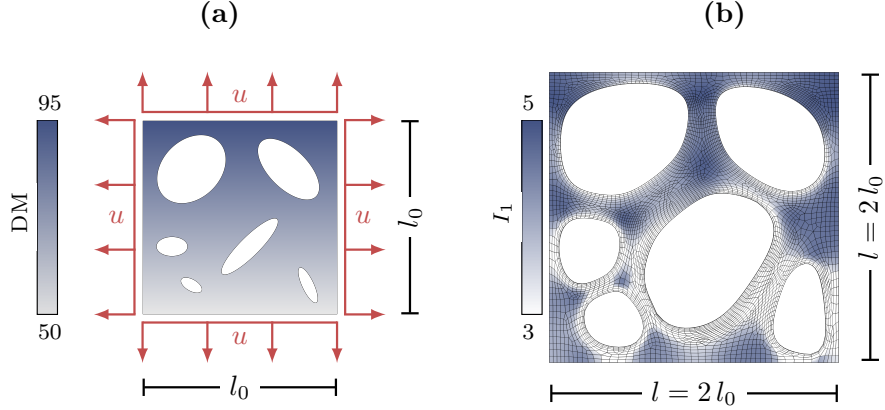
\begin{figure}[tp!]
\centering
\resizebox{0.7\textwidth}{!}{
\tikzsetnextfilename{biax_specimen}

\begin{tikzpicture}
\begin{groupplot}[
	group style = {group size = 2 by 1, vertical sep = 0.06*\textwidth,
								horizontal sep = 0.1*\textwidth},
	]
\nextgroupplot[
width=0.35*\textwidth, height=0.35*\textwidth,
ymin = 0, ymax = 9,
xmin = 0, xmax = 9,
xtick=\empty,
ytick=\empty,
xlabel shift = -14 pt,
ylabel shift = -19 pt,
hide axis,
clip=false,
title={\textbf{(a)}},
] 

\addplot graphics [xmin=2.5,xmax=7.5,ymin=2.5,ymax=7.5] {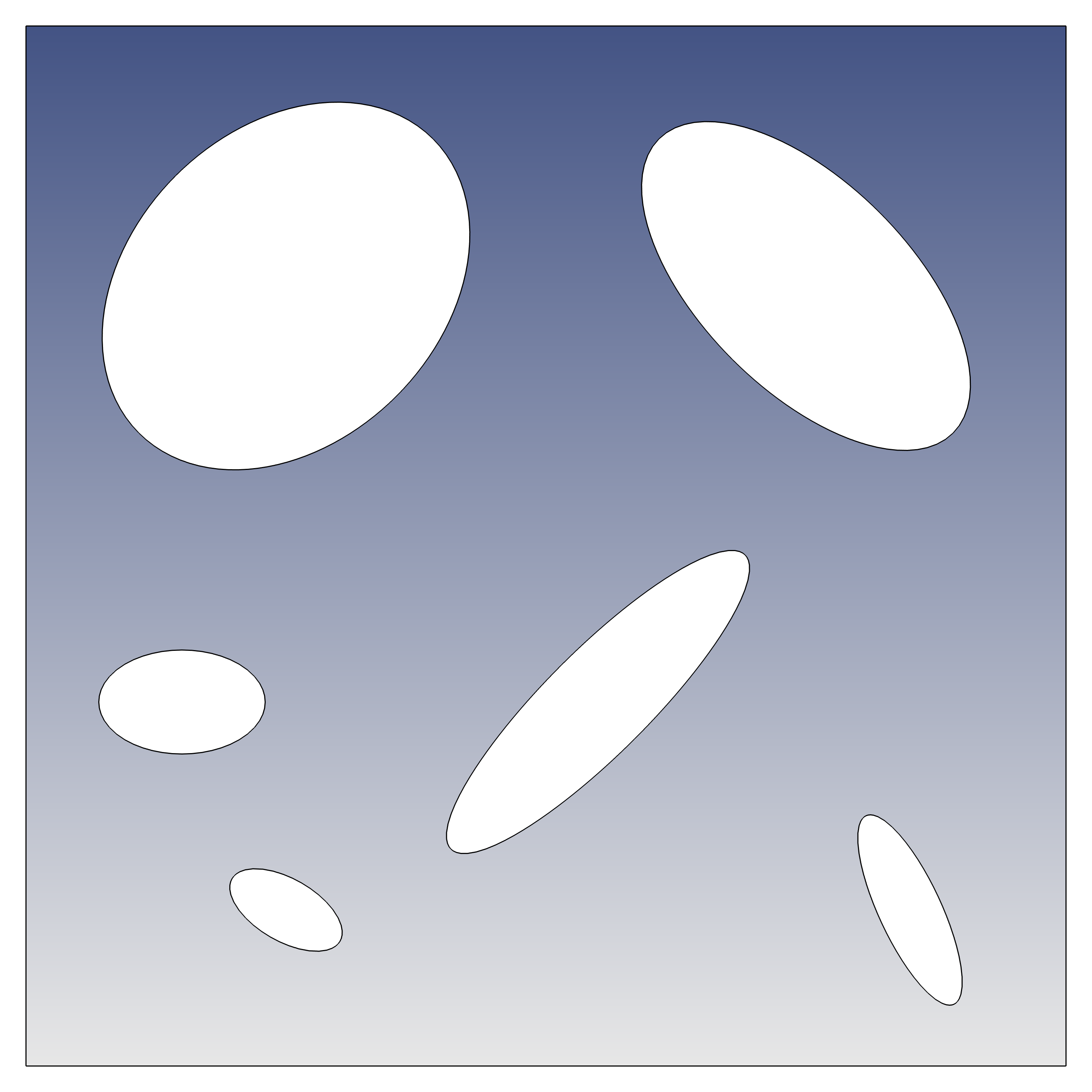};

\foreach \x in {0.00,1.6,3.2,4.8} {
    \edef\temp{\noexpand\draw[->,color=CPSred, thick,shorten <=-0.4pt] (2.6 + \x ,7.6) -- (2.6 + \x ,8.6);
    }
    \temp
}
\draw[color=CPSred,  thick] (2.6,7.6) -- (7.4,7.6);
\draw (5,7.6) node [above,CPSred] {$u$};

\foreach \x in {0.00,1.6,3.2,4.8} {
    \edef\temp{\noexpand\draw[->,color=CPSred, thick,shorten <=-0.4pt] (2.6 + \x ,2.4) -- (2.6 + \x ,1.4);
    }
    \temp
}
\draw[color=CPSred,  thick] (2.6,2.4) -- (7.4,2.4);
\draw (5,2.4) node [below,CPSred] {$u$};

\foreach \x in {0.00,1.6,3.2,4.8} {
    \edef\temp{\noexpand\draw[->,color=CPSred, thick,shorten <=-0.4pt] (2.4 ,2.6 + \x) -- (1.4 ,2.6 + \x);
    }
    \temp
}
\draw[color=CPSred,  thick] (2.4,2.6) -- (2.4,7.4);
\draw (2.4,5) node [left,CPSred] {$u$};

\foreach \x in {0.00,1.6,3.2,4.8} {
    \edef\temp{\noexpand\draw[->,color=CPSred, thick,shorten <=-0.4pt] (7.6 ,2.6 + \x) -- (8.6 ,2.6 + \x);
    }
    \temp
}
\draw[color=CPSred,  thick] (7.6,2.6) -- (7.6,7.4);
\draw (7.6,5) node [right,CPSred] {$u$};

\draw[|-|,color=black, thick] (9.2 ,2.6) -- (9.2 ,7.4) node[midway,fill=white,rotate=90] {$l_0$};
\draw[|-|,color=black, thick] (2.6 ,0.8) -- (7.4 ,0.8) node[midway,fill=white] {$l_0$};

\draw [left color=MAXgrey, right color=CPSdarkblue, thin,shading=axis,shading angle=180] (0,2.6) rectangle (0.5,7.4);
\draw (0,5) node [fill=white,rotate=90,above] {{ \footnotesize DM}};
\draw (0.2,7.4) node [fill=white,above] {{ \footnotesize 95}};
\draw (0.2,2.6) node [fill=white,below] {{ \footnotesize 50}};

\nextgroupplot[
width=0.35*\textwidth, height=0.35*\textwidth,
ymin = 0, ymax = 9,
xmin = 0, xmax = 9,
xtick=\empty,
ytick=\empty,
xlabel shift = -14 pt,
ylabel shift = -19 pt,
hide axis,
clip=false,
title={\textbf{(b)}},
] 

\addplot graphics [xmin=0,xmax=7.5,ymin=1.25,ymax=8.75] {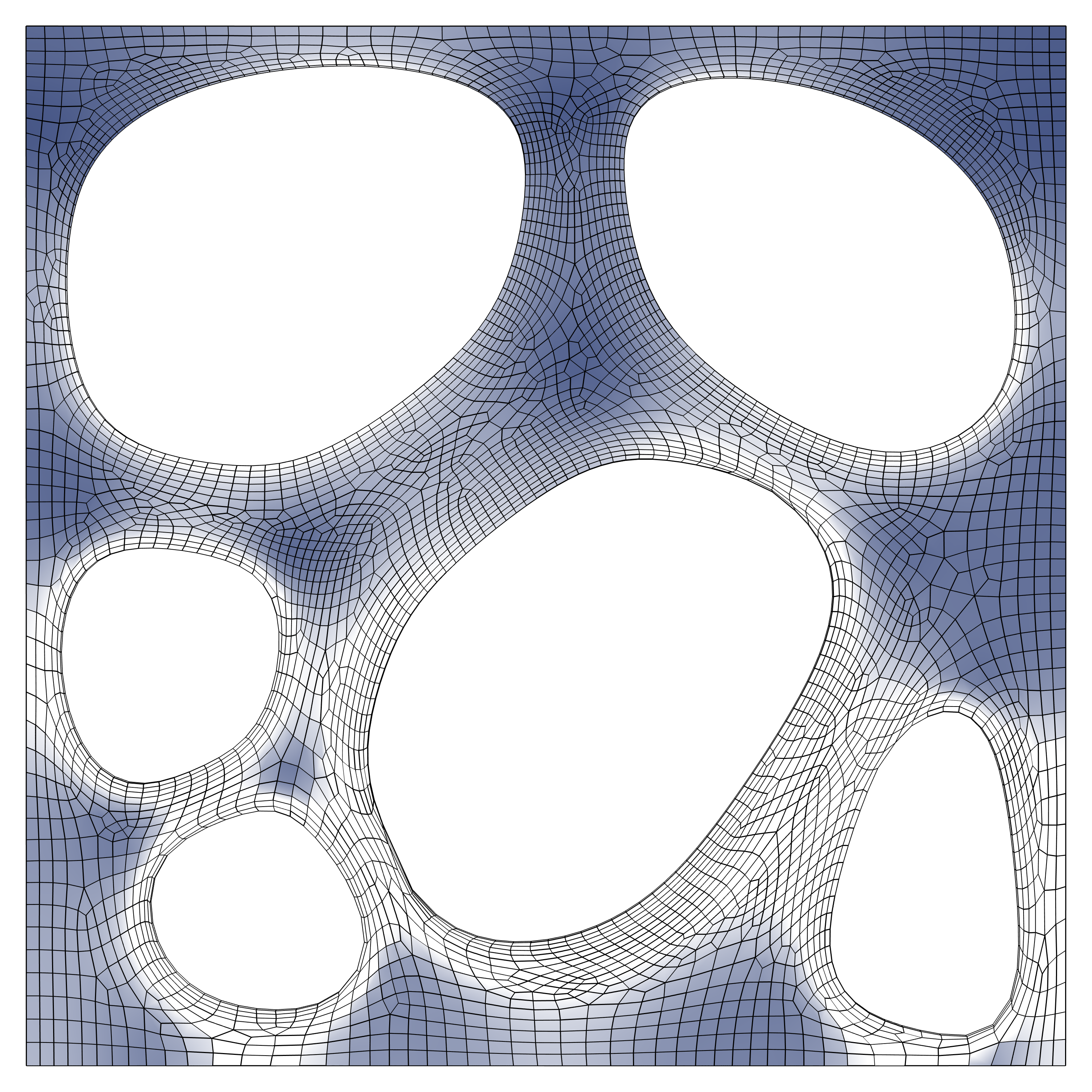};

\draw[|-|,color=black, thick] (10.6-2.5 ,0.2+1.25) -- (10.6-2.5 ,9.8-1.25) node[midway,fill=white,rotate=90] {$l=2\,l_0$};
\draw[|-|,color=black, thick] (0.2 ,-0.6+1.25) -- (9.8-2.5 ,-0.6+1.25) node[midway,fill=white] {$l=2\,l_0$};

\draw [left color=white, right color=CPSdarkblue, thin,shading=axis,shading angle=180] (-0.5,2.6) rectangle (0,7.4);
\draw (-0.5,5) node [fill=white,rotate=90,above] {{ \footnotesize $I_1$}};
\draw (-0.45,7.4) node [fill=white,above] {{ \footnotesize 5}};
\draw (-0.45,2.6) node [fill=white,below] {{ \footnotesize 3}};

\end{groupplot}

%
%

\end{tikzpicture}
}
\caption{Simulation of a biaxial tensile test with an inhomogeneous specimen. \textbf{(a):} Within the specimen plane, the displacement $u$ is applied, while the specimen can deform freely in the orthogonal direction. The specimen has a length of $l_0=1\text{cm}$ and a thickness of $0.04\text{cm}$. \textbf{(b):} Simulation result for a monotonic PANN calibrated to the DM data.}
\label{fig:FEA_biax}
\end{figure}

\paragraph{Finite-element analysis of a biaxial tensile test:} Next, we demonstrate the applicability of the monotonic PANN constitutive model in FEA including very general deformation modes. For that, we simulate a biaxial tensile test with an inhomogeneous specimen with a linear grading between DM50 and DM95 (cf.~\cref{fig:FEA_biax}(a)). Due to the holes within the specimen, the applied displacement results in very general deformation modes. 
The simulation result for the best monotonic PANN is visualized in~\cref{fig:FEA_biax}(b). Due to the grading of the structure, there is a pronounced change of shape of the holes within the specimen. In \cref{fig:mat_stab_1}(b), the resulting tensile force is visualized up to the point where the simulation does not converge anymore. The tensile force shows a similar qualitative behavior as the uniaxial tensile tests of the DM material (cf.~\cref{fig:stress_inv_1a}), with a distinct change of slope around small displacement values. For all monotonic PANN instances, the simulation almost converges until the end. For the unrestricted PANN, the simulations only converge for moderate displacements, where most simulations stop converging at around 20\% of the maximum displacement. 
Furthermore, in \cref{fig:mat_stab_1}(b), the deformation modes of the simulations are visualized for all investigated models. Since the constitutitve models depend on two strain invariants rather than the deformation gradient directly, it is sufficient to visualize the deformation modes in the invariant plane of $\Bar{I}_1$ and $\Bar{I}_2$. In this invariant plane, all deformation modes fulfilling $\det\bF=1$ form a cone \cite{Baaser_Hopmann_Schobel_2013}. The calibration data only contains uniaxial tension, which forms the lower bound of this cone. Thus, most of the deformation modes in the simulation are an extrapolation towards more general deformation modes. For the monotonic PANN, the simulations converge within a wide range of the invariant plane. In contrast, the simulations with the unrestricted PANN cover a considerably smaller range of deformations. However, this does not mean that the unrestricted PANN is not elliptic for some moderate extrapolations away from the calibration data at all. Rather, at some Gauss points in the finite element simulation, the body might experience a peak in the deformation, resulting in a loss of ellipticity at this point and, thus, non-convergence of the overall simulation.

\paragraph{Numerical investigation of the acoustic tensor:} Finally, we investigate the material stability of the PANN constitutive models by numerically evaluating the ellipticity condition \cref{eq:ellip_conds}. For the test vector $\bB$, we chose a spherical parametrization according to \textcite[Eq.~68]{klein2024a}. We investigate material stability within two areas in the invariant plane (cf.~\cref{fig:mat_stab_1}(c)), with area A containing deformations relatively close to the calibration data, and area B containing larger extrapolations away from the calibration data. The models are evaluated for multiple parameter values between DM50 and DM95.
In \cref{fig:mat_stab_1}(c), the number of elliptic (or materially stable) points is visualized for the different PANN models. All monotonic PANNs are elliptic within area A. For larger Shore A values, some of the monotonic PANNs have a loss of ellipticity for some points within area B. For the unrestricted PANN, several models show a pronounced loss of ellipticity throughout all Shore A values, even within area A. Some unrestricted PANNs lose ellipticity for almost all of the deformation modes in area B. 

\paragraph{Discussion:} Although only being calibrated on uniaxial tensile data, the monotonic PANN is applicable in complex FEA. This includes numerically challenging buckling phenomena and very general deformation modes. This is possible as the monotonic PANN learns to be materially stable in a wide range of deformations outside the calibration data. This is in contrast to the unrestricted PANN, which performs poorly in the FEA and shows a pronounced loss of material stability for extrapolation away from the calibration data. This demonstrates how monotonicity can improve the model performance in terms of material stability in multiaxial deformation scenarios.

\section{Conclusion}\label{sec:conc}

In the present work, we apply physics-augmented neural network (PANN) constitutive models to experimental uniaxial tensile data of different rubber-like materials whose behavior depends on manufacturing parameters. For this, we conduct experimental investigations on a 3D printed digital material at different mix ratios and consider several datasets from literature \cite{Liao_Hossain_Yao_2020,zhang2024,Slesarenko_Rudykh_2018}.
We introduce a parametrized hyperelastic PANN model which can represent the material behavior at different manufacturing parameters. The proposed model fulfills common mechanical conditions of hyperelasticity. In addition, the strain energy potential of the proposed model is a monotonic function in isotropic isochoric strain invariants of the right Cauchy-Green tensor. We show that, in incompressible hyperelasticity, this is a relaxed version of the ellipticity (or rank-one convexity) condition. Using this {relaxed} ellipticity condition, the PANN model remains flexible enough to be applicable to a wide range of materials while having enough structure for a stable extrapolation outside the calibration data.

In all cases studied, the proposed model shows an excellent performance. Notably, one single constitutive modeling approach performs excellently on all datasets, although they show a largely varying qualitative and quantitative stress behavior. We demonstrate how monotonicity can improve the model performance outside the calibration data. Although only calibrated on uniaxial tensile data, the monotonic PANN is applicable in complex FEA including numerically challenging buckling phenomena and very general deformation modes.
The findings of our work suggest that monotonicity could play a key role in the formulation of very general yet robust and stable constitutive models applicable to materials with highly nonlinear and parametrized behavior. Thus, our modeling framework could serve as a basis for the formulation of more sophisticated models including rate-dependent \cite{abdolazizi2024,rosenkranz2024,Holthusen_Lamm_Brepols_Reese_Kuhl_2024}, inelastic \cite{wollner2023,zlatic2024,MEYER2023105416,boes2024}, and multiphysical effects \cite{klein2022a,Fuhg_Jadoon_Weeger_Seidl_Jones_2024,kalina2024a}.


\vspace*{3ex}
{\small
\noindent
\textbf{CRediT authorship contribution statement.}  
{D.~K.~Klein:} Conceptualization, Formal analysis, Investigation, Methodology, Visualization, Software, Validation, Writing -- original draft, Writing -- review \& editing, Funding acquisition. {M.~Hossain:} Conceptualization, Methodology, Writing -- review \& editing, Funding acquisition, Resources. {M.~Kannapin:} Investigation, Visualization, Methodology, Software, Validation. {S.~Rudykh:} Investigation, Resources, Data Curation. {K.~Kikinov:} Investigation, Data Curation. {A.~J.~Gil:} Conceptualization, Methodology, Writing -- review \& editing, Funding acquisition, Resources.

\vspace*{0.5ex}

\noindent
\textbf{Acknowledgment.}  
D.K.~Klein acknowledges funding from the Deutsche Forschungsgemeinschaft (DFG, German Research Foundation, project number 492770117) and support by a fellowship of the German Academic Exchange Service (DAAD).
D.K.~Klein and M.~Kannapinn acknowledge support by the Graduate School of Computational Engineering at TU Darmstadt. 
M. Hossain  acknowledges the support of the EPSRC Impact Acceleration Account (EP/X525637/1) and the Royal Society (UK) through the International Exchange Grant (IEC/NSFC/211316).
S. Rudykh and K. Kikinov acknowledge support of the European Research Council (ERC) through Grant No. 852281 -- MAGIC.
A.J. Gil acknowledges the financial support provided by UK Defence, Science and Technology Laboratory through grant DSTLX 10000157545 and The Leverhulme Trust.
\vspace*{0.5ex}

\noindent
\textbf{Declaration of competing interest.} The authors declare that they have no conflict of interest.
}

 
\appendix
\numberwithin{equation}{section}

\renewcommand*{\bibfont}{\small}
\printbibliography

\end{document}